\documentstyle[12pt]{article}
\def\mbar(#1){\overline #1 \hskip 1pt{}}

\def\der{\partial}

\def\derover(#1,#2){ { \der_{#1} \over \der_{#2} } }
\def\qq(#1){{\hskip 2pt\hbox{\bm #1}\hskip 2pt}} 
\def\sp(#1){ \noalign{\vskip #1pt} }

\def\slash(#1){ { {#1} \hskip -8.3pt \big/ } }
\def\Slash(#1){ { {#1} \hskip -9.6pt \big/ } }

\newcommand{\cropen}{\crcr\noalign{\vskip 5pt}}
\newcommand{\Ctilde}{{\widetilde C}}
\newcommand{\Dtilde}{{\widetilde D}}
\newcommand{\Vtilde}{{\widetilde V}}
\newcommand{\Cbar}{{\bar C}}

\newcommand{\psnu}{\partial_\nu}

\newcommand{\pnu}{\partial^\nu}

\newcommand{\Gh}{\hbox{Gh}}
\newcommand{\ie}{{\it i.e.}}

\renewcommand{\theequation}{\arabic {section}.\arabic{equation}}

\makeatletter
\def\eqnarray{%
 \stepcounter{equation}%
 \let\@currentlabel=\theequation
 \global\@eqnswtrue
 \global\@eqcnt\z@
 \tabskip\@centering
 \let\\=\@eqncr
 $$\halign to \displaywidth\bgroup\@eqnsel\hskip\@centering
 $\displaystyle\tabskip\z@{##}$&\global\@eqcnt\@ne
 \hfil$\displaystyle{{}##{}}$\hfil
 &\global\@eqcnt\tw@$\displaystyle\tabskip\z@{##}$\hfil
 \tabskip\@centering&\llap{##}\tabskip\z@\cr}
\makeatother

\makeatletter
\def\@arrayacol{\edef\@preamble{\@preamble \hskip .5\arraycolsep}}
\def\array{\let\@acol\@arrayacol \let\@classz\@arrayclassz
 \let\@classiv\@arrayclassiv \let\\\@arraycr\def\@halignto{}\@tabarray}
\makeatother
\renewcommand{\arraystretch}{1.6}

\begin{document}

\setlength{\baselineskip}{7mm}
\begin{titlepage}
\begin{flushright}
EPHOU-98-003 \\
April, 1998
\end{flushright}
 
%\vspace{30mm}

\vspace{15mm}
\begin{center} 
{\Large Quantization of Even-Dimensional Actions  \\}
{\Large of Chern-Simons Form with Infinite Reducibility}\\
\vspace{1cm}
{\bf {\sc Noboru Kawamoto, Kazuhiko Suehiro, Takuya Tsukioka\\
and Hiroshi Umetsu\footnote{
Address after April 1998: Dept. of Phys. Osaka Univ. Toyonaka, 
Osaka 560-0043, Japan\\
${}$\qquad umetsu@phys.wani.osaka-u.ac.jp}}}\\
{\it{ Department of Physics, Hokkaido University }}\\
{\it{ Sapporo, 060-0810, Japan}}\\
{kawamoto, suehiro, tsukioka, umetsu@particle.sci.hokudai.ac.jp}
\end{center}
\vspace{2cm}

\begin{abstract}
We investigate the quantization of even-dimensional topological 
actions of Chern-Simons form which were proposed previously.
We quantize the actions by Lagrangian and Hamiltonian formulations 
{\`a} la Batalin, Fradkin and Vilkovisky. 
The models turn out to be infinitely reducible and thus we need 
an infinite number of ghosts and antighosts.
The minimal actions of Lagrangian formulation 
which satisfy the master equation of 
Batalin and Vilkovisky have the same Chern-Simons form 
as the starting classical actions.
In the Hamiltonian formulation we have used the formulation 
of cohomological perturbation and explicitly shown that 
the gauge-fixed actions of both formulations coincide 
even though the classical action breaks Dirac's regularity 
condition.
We find an interesting relation that the BRST charge of Hamiltonian 
formulation is the odd-dimensional fermionic counterpart of the topological 
action of Chern-Simons form. 
Although the quantization of two dimensional models which include 
both bosonic and fermionic gauge fields are investigated in detail,
it is straightforward to extend the quantization into arbitrary 
even dimensions.
This completes the quantization of previously proposed topological gravities 
in two and four dimensions. 
\end{abstract}

\end{titlepage}

%%%%%%%%%%%%%%%%%%%%
\section{Introduction}

\setcounter{equation}{0}
\setcounter{footnote}{0}

It is obviously the most challenging problem how we can formulate the quantum 
gravity and the standard model in a unifying and constructive way. 
Toward a possible solution to this problem, the current trend is heading 
to the string related topics. 
It is, however, not obvious if the string is the only possible direction 
to this problem. 
In fact two-dimensional quantum gravity was formulated by a lattice gravity, 
the dynamical triangulation of random surface, equivalently by the 
matrix model. 
On the other hand three-dimensional Einstein gravity was successfully 
formulated by the Chern-Simons action even at the quantum level~\cite{w1}. 
Since the Chern-Simons action is composed of one-form gauge field, the 
general covariance is automatic in the formulation. 

If we combine the two successful formulations to find a new 
formulation of quantum gravity, we naturally lead to an idea that 
we should find a gauge theory formulated by all degrees of differential forms. 
Here each form corresponds to a fundamental simplex of a simplicial manifold. 
There are also good reasons that gravity theory can be formulated by a gauge 
theory. 

The standard Chern-Simons action partly satisfies the above criteria. 
Previously one of the authors (N.K.) and Watabiki have proposed 
a new type of topological actions in arbitrary dimensions 
which have the Chern-Simons form~\cite{kw1}.
The actions have the same algebraic structure as the ordinary Chern-Simons 
action and are formulated by all degrees of differential forms. 
It was shown that two-dimensional topological gravities~\cite{kw2}
and a four-dimensional topological conformal gravity~\cite{kw3}
were formulated by the even-dimensional version 
of the generalized Chern-Simons actions at the classical level.
It may not be an unnatural expectation that this type of formulation 
could play an important role in the formulation of quantum gravity. 

Since the topological gravity theories mentioned above are defined 
at the classical level, 
it is natural to ask if they are well defined at the quantum level. 
In this paper we investigate the quantization of the models defined by 
the generalized Chern-Simons actions.
In the analyses we don't specify an algebra in a particular way to 
accommodate some gravity theory.
The stress should be on the quantization of 
the model itself.  
It turns out that the quantization of the generalized Chern-Simons action 
is highly nontrivial.
There are two difficulties in quantizing these models. 
Firstly the action has a zero form square term 
multiplied by the highest form and thus the vanishing condition of 
the zero form square is the equation of motion which breaks 
regularity condition~\cite{ht,kos}. 
Secondly the theory is highly reducible, in fact infinitely reducible, 
as we show later. 

In quantizing reducible systems we need to use the Lagrangian formulation 
developed by Batalin and Vilkovisky~\cite{bv}. 
In order to clarify the role of the violation of Dirac's regularity 
condition we quantize the system in the Hamiltonian formulation of 
Batalin, Fradkin and Vilkovisky~\cite{bfv}. 
In the concrete analyses the quantization procedure of cohomological 
perturbation developed by Henneaux et al.~\cite{ht} is used.  

It was shown in the quantization of the simplest abelian version of 
generalized Chern-Simons action that the particular type of regularity 
violation does not cause serious problems for the quantization~\cite{kos}. 
In this paper we investigate nonabelian version of Chern-Simons actions 
which turn out to be infinitely reducible while it was not the case 
for the abelian version. 
We show that the gauge-fixed action derived from the Lagrangian 
formulation leads to the same result of 
the Hamiltonian formulation even with regularity violating 
constraints and infinite reducibility. 

In defining gauge fields and parameters, we introduce quaternion valued 
gauge fields and parameters containing forms of all possible degrees. 
They play an important role to manipulate the quantization in a unified 
manner. 
In other words the infinite number of ghost fields can
be unified into a compact form and thus the quantization procedure can 
be largely simplified and becomes transparent even with the infinite 
reducibility. 
It is also interesting to note that the nonabelian version of the generalized 
Chern-Simons actions provide very fruitful and nontrivial examples for 
the quantization 
of infinitely reducible systems to be compared with 
Brink-Schwarz superparticle~\cite{bs}, 
Green-Schwarz superstring~\cite{gs} 
and covariant string field theories~\cite{w2}. 

This paper is organized as follows: 
We first briefly summarize the formulation of the generalized Chern-Simons 
actions in section 2. 
We then  explain the quantization of two-dimensional generalized 
Chern-Simons model in the Lagrangian formulation in section 3. 
In section 4 the quantization of the same model is carried out by 
the Hamiltonian formulation. 
Section 5 explains the analyses of the perturbative aspects of the 
two-dimensional models. 
In section 6 we extend the quantization procedure of the two-dimensional 
model into arbitrary even dimensions, in particular we discuss 
a four-dimensional model as an example. 
Conclusions and discussions are given in the final section. 

%%%%%%%%%%%%%%%%%%%%
\section{Generalization of Chern-Simons action into arbitrary dimensions}

\setcounter{equation}{0}
\setcounter{footnote}{0}

The generalized Chern-Simons actions, which were proposed by 
one of the authors (N.K.) and Watabiki some years ago, is a generalization 
of the ordinary 
three-dimensional Chern-Simons theory into arbitrary dimensions~\cite{kw1}.
We summarize the results in this section.
The essential point of the generalization is to extend a one-form gauge field 
and zero-form gauge parameter to a quaternion valued 
generalized gauge field and gauge parameter which contain forms of all 
possible degrees.
Correspondingly the standard gauge symmetry is extended to much higher 
topological symmetry.
These generalizations are formulated in such a way that the generalized 
actions have the same algebraic structure as the ordinary three-dimensional 
Chern-Simons action. 

In the most general form, a generalized gauge field $\cal{A}$ and a gauge 
parameter $\cal{V}$ are defined by the following component form:
\begin{eqnarray}
 {\cal A} & = & {\mbox{\bf 1}}\psi + {\mbox{\bf i}} \hat{\psi} +  
            {\mbox{\bf j}} A + {\mbox{\bf k}} \hat{A}, 
\label{eqn:ggf} \\
 {\cal V} & = & {\mbox{\bf 1}} \hat{a} + {\mbox{\bf i}} a +  
            {\mbox{\bf j}} \hat{\alpha} + {\mbox{\bf k}} \alpha, 
\label{eqn:ggp}
\end{eqnarray}
where $( \psi, \alpha )$, $( \hat{\psi}, \hat{\alpha} )$, 
$( A, a )$ and  $( \hat{A},\hat{a} )$ are direct sums of 
fermionic odd forms, fermionic even forms, bosonic odd forms 
and bosonic even forms, respectively, and they take values on a gauge algebra.
The bold face symbols ${\bf 1}$, ${\bf i}$,  ${\bf j}$ and ${\bf k}$ 
satisfy the algebra
\begin{equation}
\begin{array}{c}
{\bf 1}^2={\bf 1}, \quad {\bf i}^2=\epsilon_1 {\bf 1}, \quad 
{\bf j}^2=\epsilon_2 {\bf 1}, \quad {\bf k}^2=-\epsilon_1 \epsilon_2 {\bf 1},
\\
{\bf i}{\bf j}=-{\bf j}{\bf i}={\bf k}, \quad 
{\bf j}{\bf k}=-{\bf k}{\bf j}=-\epsilon_2 {\bf i}, \quad 
{\bf k}{\bf i}=-{\bf i}{\bf k}=-\epsilon_1 {\bf j},
\end{array}
\end{equation}
where $(\epsilon_1,\epsilon_2)$ takes the value $(-1,-1), (-1,+1), (+1,-1)$ or 
$(+1,+1)$.
Throughout this paper we adopt the convention 
$(\epsilon_1,\epsilon_2)=(-1,-1)$, 
then the above algebra corresponds to the quaternion algebra.
The following graded Lie algebra can be adopted as a gauge algebra:
\begin{eqnarray}
\left[ T_a , T_b \right] &=& f^c_{ab} T_c, \nonumber\\
\left[ T_a , \Sigma_\beta \right] &=& g^\gamma_{a\beta} \Sigma_\gamma, \\
\left\{ \Sigma_\alpha , \Sigma_\beta \right\} &=& h^c_{\alpha\beta} T_c, 
\nonumber
\end{eqnarray}
where all the structure constants are subject to consistency conditions 
which follow from the graded Jacobi identities.
If we choose $\Sigma_\alpha=T_a$ especially, this algebra reduces to 
$T_a T_b = k^c_{ab}T_c$ which is closed under multiplication.
A specific example of such algebra is realized by Clifford algebra~\cite{kw2}.
The components of the gauge field $\cal{A}$ and parameter $\cal{V}$ are 
assigned to the elements of the gauge algebra in a specific way:
\begin{equation}
 \begin{array}{rclrclrclrcl}
 A          &=& T_a A^a, \quad 
 \hat{\psi} &=& T_a \hat{\psi}^a, \quad 
 \psi       &=& \Sigma_\alpha \psi^\alpha, \quad 
 \hat{A}    &=& \Sigma_\alpha \hat{A}^\alpha, \\ 
 \hat{a}    &=& T_a \hat{a}^a, \quad
 \alpha     &=& T_a \alpha^a, \quad
 \hat{\alpha}&=&\Sigma_\alpha \hat{\alpha}^\alpha, \quad 
 a          &=& \Sigma_\alpha a^\alpha.
 \end{array}
\end{equation}
An element having the same type of component expansion as $\cal{A}$ 
and $\cal{V}$ belong to $\Lambda_-$ and $\Lambda_+$ class, respectively, and 
these elements fulfill 
the following $Z_2$ grading structure:
$$
[\lambda_+ , \lambda_+] \in \Lambda_+, \qquad
[\lambda_+ , \lambda_-] \in \Lambda_-, \qquad
\{\lambda_- , \lambda_-\} \in \Lambda_+, 
$$
where $\lambda_+ \in \Lambda_+$ and $\lambda_- \in \Lambda_-$.
The elements of $\Lambda_-$ and $\Lambda_+$ can be regarded as generalizations 
of odd forms and even forms, respectively.
In particular the generalized exterior derivative which belongs to 
$\Lambda_-$ is given by
\begin{equation}
Q={\bf j}d, 
\end{equation}
and the following relations similar to the ordinary differential algebra hold:
$$
\{Q , \lambda_-\}=Q\lambda_-, \qquad
[Q , \lambda_+]=Q\lambda_+, \qquad
Q^2=0,
$$
where $\lambda_+ \in \Lambda_+$ and $\lambda_- \in \Lambda_-$.
To construct the generalized Chern-Simons actions, 
we need to introduce the two kinds of traces 
\begin{equation}
 \begin{array}{rcl}
 \mbox{Htr}[T_a , \cdots]=0, 
   &{\qquad}& \mbox{Htr}[\Sigma_\alpha , \cdots]=0, \\
 \mbox{Str}[T_a , \cdots]=0, 
   &{\qquad}& \mbox{Str}\{\Sigma_\alpha , \cdots\}=0, \\  
 \end{array}                                                
\label{eqn:tr}
\end{equation}
where $(\cdots)$ in the commutators or the anticommutators denotes a product 
of generators. In particular $(\cdots)$ should include an odd number
of $\Sigma_\alpha$'s in the last eq. of (\ref{eqn:tr}).
These definitions of the traces are crucial 
to show that the generalized Chern-Simons action can be invariant under 
the generalized gauge transformation presented bellow.

After the above preparations, we can define four types of actions which have 
Chern-Simons form, 
\begin{equation}
 \begin{array}{ccc}
 \displaystyle{S^b_e=\int_M \mbox{Htr}_{\bf k}
        \left( \frac{1}{2}{\cal A}Q{\cal A}+\frac{1}{3}{\cal A}^3 \right)},
         &{\quad}&
 \displaystyle{S^f_o=\int_M \mbox{Htr}_{\bf 1}
        \left( \frac{1}{2}{\cal A}Q{\cal A}+\frac{1}{3}{\cal A}^3 \right)},\\
 \displaystyle{S^b_o=\int_M \mbox{Str}_{\bf j}
        \left( \frac{1}{2}{\cal A}Q{\cal A}+\frac{1}{3}{\cal A}^3 \right)}, 
         &{\quad}&
 \displaystyle{S^f_e=\int_M \mbox{Str}_{\bf i}
        \left( \frac{1}{2}{\cal A}Q{\cal A}+\frac{1}{3}{\cal A}^3 \right)},
\label{eqn:gCSa}
                                                     \raisebox{8mm}{}     
 \end{array}
\end{equation}
where $S^b_e, S^f_o, S^b_o$ and $S^f_e$ are an even-dimensional bosonic 
action, an odd-dimensional fermionic action, an odd-dimensional bosonic action 
and an even-dimensional fermionic action, respectively.
$\mbox{Htr}_{\bf q}(\cdots)$ and $\mbox{Str}_{\bf q}(\cdots) \ \ 
({\bf q}={\bf 1}, {\bf i}, {\bf j}, {\bf k})$ are defined so as to pick up 
only the coefficients of ${\bf q}$ from $(\cdots)$ and take the traces defined 
by eq.(\ref{eqn:tr}).
The reason why we obtain the four different types of action is related to 
the fact that the Chern-Simons term in the trace, 
$\frac{1}{2}{\cal A}Q{\cal A}+\frac{1}{3}{\cal A}^3 $, 
belongs to $\Lambda_-$ class and thus possesses the 
four different component types, the same types as in ${\cal A}$ of 
(\ref{eqn:ggf}).  
We then need to pick up $d$-form terms to obtain $d$ dimensional actions 
defined on a manifold $M$.
These actions are invariant up to surface terms under the 
generalized gauge transformation
\begin{equation}
\delta {\cal A}=[Q+{\cal A} , {\cal V} ], 
\label{eqn:ggt}
\end{equation}
where ${\cal V}$ is the generalized gauge parameter defined by 
eq.(\ref{eqn:ggp}).
It should be noted that this symmetry is much larger than the usual gauge 
symmetry, in fact topological symmetry, since the gauge parameter ${\cal V}$ 
contains as many gauge parameters as gauge fields of various forms.

Substituting eqs.(\ref{eqn:ggf}) and (\ref{eqn:ggp}) into eqs.(\ref{eqn:gCSa}) 
and (\ref{eqn:ggt}), we obtain explicit forms of the actions
\begin{eqnarray}
S^b_e &=& - \int_M \mbox{Htr}
   \left\{ \hat{A}(dA+A^2+\hat{\psi}^2-\psi^2)+\frac{1}{3}\hat{A}^3
             +\psi(d\hat{\psi}+[A,\hat{\psi}]) \right\}, \nonumber \\
S^f_o &=& - \int_M \mbox{Htr}
   \left\{ \psi(dA+A^2+\hat{\psi}^2+\hat{A}^2)-\frac{1}{3}\psi^3
             -\hat{A}(d\hat{\psi}+[A,\hat{\psi}]) \right\}, \nonumber \\
S^b_o &=& - \int_M \mbox{Str}
   \left\{ \frac{1}{2}AdA+\frac{1}{3}A^3
             -\frac{1}{2}\hat{\psi}(d\hat{\psi}+[A,\hat{\psi}])
                               + \hat{\psi} \{\psi,\hat{A}\} \right.
\label{eqn:comexp} \\
      & & \left. \hspace{3cm}
          -\frac{1}{2}\psi(d\psi+\{A,\psi\})
          -\frac{1}{2}\hat{A}(d\hat{A}+[A,\hat{A}]) \right\}, \nonumber \\
S^f_e &=& - \int_M \mbox{Str}
   \left\{ \hat{\psi}(dA+A^2-\psi^2+\hat{A}^2)
          +\frac{1}{3}\hat{\psi}^3+\hat{A}(d\psi+\{A,\psi\}) \right\},\nonumber
\end{eqnarray}
and the gauge transformations
\begin{equation}
 \begin{array}{rcl}
  \delta A 
   &=& d\hat{a}+[A,\hat{a}]-\{\hat{\psi},\alpha\}
       +[\psi,\hat{\alpha}]+\{\hat{A},a\}, \\
  \delta \hat{\psi} 
   &=& d\alpha+\{A,\alpha\}+[\hat{\psi},\hat{a}]
       +[\psi,a]-\{\hat{A},\hat{\alpha}\}, \\
  \delta \psi 
   &=& -d\hat{\alpha}-[A,\hat{\alpha}]-[\hat{\psi},a]
        +[\psi,\hat{a}]-[\hat{A},\alpha], \\
  \delta \hat{A} 
   &=& -da-\{A,a\}+\{\hat{\psi},\hat{\alpha}\}
       +[\psi,\alpha]+[\hat{A},\hat{a}],               
\label{eqn:gt-com} 
 \end{array} 
\end{equation}
where $[ \ , \ ]$ and $\{ \ , \ \}$ are commutator and anticommutator, 
respectively. 

Each action in (\ref{eqn:gCSa}) leads to the same equation of motion
\begin{equation}
{\cal F}=Q{\cal A}+{\cal A}^2=0, 
\label{eqn:f-curv}
\end{equation}
where ${\cal F}$ is a generalized curvature and thus the equation of motion 
is a vanishing curvature condition.
Component expansions of the equations of motion are given by
\begin{equation}
 \begin{array}{rcl}
  dA+A^2+\hat{\psi}^2-\psi^2+\hat{A} &=& 0, \\
  d\hat{\psi}+[A,\hat{\psi}]-\{\psi,\hat{A}\} &=& 0, \\
  d\psi+\{A,\psi\}+[\hat{A},\hat{\psi}] &=& 0, \\
  d\hat{A}+[A,\hat{A}]+\{\hat{\psi},\psi\} &=& 0.     
\label{eqn:eqmotions}
 \end{array}
\end{equation}

We now show the explicit forms of two- and four-dimensional actions which 
will be used in this paper.
We introduce the following notations: 
\begin{eqnarray}
 {\cal A} & = & {\mbox{\bf 1}}\psi + {\mbox{\bf i}} \hat{\psi} +  
            {\mbox{\bf j}} A + {\mbox{\bf k}} \hat{A} \nonumber \\
          & \equiv & {\mbox{\bf 1}}(\chi_1+\chi_3) + 
                     {\mbox{\bf i}} (\chi_0+\chi_2+\chi_4) +  
                     {\mbox{\bf j}} (\omega+\Omega) + 
                     {\mbox{\bf k}} (\phi+B+H),             
\label{eqn:ggff} 
\end{eqnarray}
where $\phi$, $\omega$, $B$, $\Omega$, $H$ are bosonic 0-, 1-, 2-, 3-, 4-form, 
and $\chi_0$, $\chi_1$, $\chi_2$, $\chi_3$, $\chi_4$ are 
fermionic 0-, 1-, 2-, 3-, 4-form, respectively.
Substituting the above expressions 
into bosonic even action $S^b_e$ of (\ref{eqn:comexp}) and taking the 
two-form part, 
we obtain the two-dimensional generalized Chern-Simons action
\begin{equation}
         S_2=-\int{\mbox{Htr}}
               \Bigl\{
                      \phi(d\omega+\omega^2+\{\chi_0,\chi_2 \}-\chi_1^2)
                      +B(\phi^2+\chi_0^2)
                      +\chi_1(d\chi_0+[\omega,\chi_0])
               \Bigr\}. 
\label{eqn:S2} 
\end{equation}
Similarly the four-dimensional generalized Chern-Simons action can 
be obtained 
by taking the four-form part of the bosonic even action $S^b_e$: 
\begin{equation}
         S_4=-\int{\mbox{Htr}}
               \Bigl\{
                      B(d\omega+\omega^2)
                      +\phi(d\Omega+\{\omega,\Omega\}+B^2) 
                      +\phi^2H 
               \Bigr\}, 
\label{eqn:4d-act} 
\end{equation}
where we have omitted fermions for simplicity ($\psi=\hat{\psi}=0$). 
Component wise explicit forms of gauge transformations and equations 
of motions for these actions can be obtained 
by eqs. (\ref{eqn:gt-com}) and (\ref{eqn:eqmotions}) respectively 
and will be given later.

%%%%%%%%%%%%%%%%%%%%
\section{Lagrangian quantization of two-dimensional models} 

\setcounter{equation}{0}
\setcounter{footnote}{0}

\subsection{Infinite reducibility}

Hereafter we consider the even-dimensional bosonic action $S^b_e$ of 
(\ref{eqn:comexp}), in particular the two-dimensional version 
(\ref{eqn:S2}) 
with a nonabelian gauge algebra 
as a concrete example although we will see that models 
in arbitrary even dimensions can be treated in the similar way.
A simple example for nonabelian gauge algebras 
is given by Clifford algebra $c(0,3)$ generated 
by $\{T_a\}=\{1,i\sigma_k; k=1,2,3\}$ where $\sigma_k$'s are 
Pauli matrices~\cite{kw2}. 
Here we choose $\Sigma_\alpha=T_a$ for simplicity and thus the algebra 
is closed under multiplication.
In this case the Htr satisfying the conditions in (\ref{eqn:tr}) reduces 
to the normal trace for matrices; Htr $\rightarrow$ Tr. 

The quantization of a purely bosonic model, which is obtained by omitting 
fermionic gauge fields and parameters 
in the classical action and gauge transformations, was investigated
in the previous paper~\cite{kstu1}.
Here we keep fermionic fields and thus investigate the most general model
in two dimensions.
Then the action expanded into components is given by 
\begin{eqnarray}
 S_{0}=-\int d^2x{\mbox{Tr}}\epsilon^{\mu\nu}
        \Bigl\{\hspace*{-1mm}
               &\phi& \hspace*{-1mm}
                      (\partial_{\mu}\omega_{\nu}+\omega_{\mu}\omega_{\nu}
                       +\hbox{$\frac{1}{2}$}\{\chi,\chi_{\mu\nu}\}
                       -\chi_\mu\chi_\nu) \nonumber \\
               &+& \hspace*{-1mm}
                   \hbox{$\frac{1}{2}$}B_{\mu\nu}(\phi^2+\chi^2)
                   +\chi_\mu(\partial_\nu\chi+[\,\omega_\nu,\chi])\Bigr\},
\label {eqn:ba}
\end{eqnarray}
where $\epsilon^{01}=1$ and $\phi$, $\omega_\mu$ and $B_{\mu\nu}$ are 
scalar, vector and antisymmetric tensor fields while 
$\chi$, $\chi_\mu$ and $\chi_{\mu\nu}$ are fermionic fields of scalar,
vector and antisymmetric tensor, respectively\rlap.
\footnote{
%%%%%%%%%footnote%%%%%%%%%%%
Throughout this paper we impose 
$\phi^{\dagger}=-\phi$, $\omega_{\mu}^{\dagger}=-\omega_{\mu}$, 
$B^{\dagger}_{\mu\nu}=B_{\mu\nu}$, $\chi^{\dagger}=-\chi$, 
$\chi_{\mu}^{\dagger}=\chi_{\mu}$ and  
$\chi^{\dagger}_{\mu\nu}=\chi_{\mu\nu}$ 
to make the classical action hermitian.
%%%%%%%%%%%%%%%%%%%%%%%%%%%%
}
This Lagrangian possesses the following gauge symmetries 
corresponding to eq.(\ref{eqn:gt-com}):
\begin{equation}
 \begin{array}{rcl}
  \delta\phi &=& [\phi, v_1]-\{\chi, \xi_1\}, \\
  \delta\omega_\mu &=& \partial_\mu v_1+[\omega_{\mu}, v_1]-\{\phi, u_{1\mu}\}
                       -[\chi_\mu, \xi_1]-\{\chi, \xi_{1\mu}\}, \\
  \delta B &=& \epsilon^{\mu\nu}(\partial_\mu u_{1\nu}+[\omega_\mu, u_{1\nu}]) 
               +[B, v_1]+[\phi, b_1] \\
           & & +\epsilon^{\mu\nu}\{\chi_\mu, \xi_{1\nu}\}
               -\{\tilde{\chi}, \xi_1\} -\{\chi, \tilde{\xi}_1\}, \\
  \delta\chi &=& \{\phi, \xi_1\}+[\chi, v_1],  \\
  \delta\chi_\mu &=& \partial_\mu\xi_1+[\omega_{\mu}, \xi_1]
                     -[\phi, \xi_{1\mu}]+[\chi_\mu, v_1]+[\chi, u_{1\mu}], \\
  \delta\tilde{\chi} &=& \epsilon^{\mu\nu}
                         (\partial_\mu\xi_{1\nu}+[\omega_\mu, \xi_{1\nu}]) 
                         +\{B, \xi_1\}+\{\phi, \tilde{\xi}_1\} \\
                     & & -\epsilon^{\mu\nu}[\chi_\mu, u_{1\nu}]
                         +[\tilde{\chi}, v_1]+[\chi, b_1], 
\label{eqn:dchi}
 \end{array}
\end{equation}
where $B$, $b_1$, $\tilde{\chi}$ and $\tilde{\xi}_1$ are defined 
by $B\equiv\frac{1}{2}\epsilon^{\mu\nu}B_{\mu\nu}$,
$b_1\equiv\frac{1}{2}\epsilon^{\mu\nu}b_{1\mu\nu}$,
$\tilde{\chi}\equiv\frac{1}{2}\epsilon^{\mu\nu}\chi_{\mu\nu}$
and $\tilde{\xi}_1\equiv\frac{1}{2}\epsilon^{\mu\nu}\xi_{1\mu\nu}$, 
respectively.
For the later convenience we put the suffix 1 for the gauge parameters.
Equations of motion (\ref{eqn:eqmotions}) for this model are given by 
\begin{eqnarray}
 \phi: & & \quad -\epsilon^{\mu\nu} 
                 (\partial_{\mu}\omega_{\nu}+\omega_{\mu}\omega_{\nu} 
                  -\chi_\mu\chi_\nu)
                 -\{\phi, B\}-\{\chi, \tilde{\chi}\}=0, \nonumber\\
 \omega_{\mu}: & & \quad  -\epsilon^{\mu\nu} 
                          (\partial_{\nu}\phi+[\omega_{\nu}, \phi] 
                          +\{\chi_\nu, \chi\})=0,  \nonumber\\  
 B: & & \quad -(\phi^2+\chi^2)=0, \nonumber \\  
 \chi: & & \quad -\epsilon^{\mu\nu}(\partial_\mu\chi_{\nu} 
                  +[\omega_\mu, \chi_{\nu}]) 
                  -[\phi, \tilde{\chi}]-[B, \chi]=0, \\
 \chi_\mu: & & \quad \epsilon^{\mu\nu}(\partial_\nu\chi 
                     +[\omega_{\nu}, \chi]
                     -\{\phi, \chi_\nu\})=0, \nonumber\\
 \tilde{\chi}: & & \quad -[\phi, \chi]=0, \nonumber
\end{eqnarray}
where we have taken the right derivative of the corresponding fields.

As in the case of the purely bosonic model~\cite{kstu1}, 
this system is infinitely on-shell reducible.
The reducibility is easily shown by extending the proof
given for the purely bosonic model.
First we introduce generalized variables
\begin{eqnarray}
 {\cal{V}}_{2n} &=& 
              {\mbox{\bf 1}} \xi_{2n \mu}dx^{\mu} 
              + {\mbox{\bf i}} \bigg( \xi_{2n}+\frac{1}{2} \xi_{2n \mu \nu} 
                    dx^{\mu} \wedge dx^{\nu} \bigg) \nonumber \\
              & & 
              +{\mbox{\bf j}} u_{2n \mu}dx^{\mu} 
              + {\mbox{\bf k}} \bigg( v_{2n} + \frac{1}{2} b_{2n \mu \nu} 
                        dx^{\mu} \wedge dx^{\nu} \bigg) \ \in \Lambda_{-}, 
\label{eqn:Ve}\\
 {\cal{V}}_{2n+1} & = & 
               {\mbox{\bf 1}} \bigg( v_{2n+1} + 
                \frac{1}{2} b_{2n+1 \mu \nu} dx^{\mu} \wedge dx^{\nu} \bigg)  
              - {\mbox{\bf i}}\,  u_{2n+1 \mu}dx^{\mu} \nonumber \\
             & & 
               -{\mbox{\bf j}} \bigg( \xi_{2n+1} + 
                \frac{1}{2}\xi_{2n+1 \mu\nu} dx^{\mu}\wedge dx^{\nu} \bigg)  
               + {\mbox{\bf k}}\, \xi_{2n+1 \mu}dx^{\mu} \  \in \Lambda_{+}, 
\label{eqn:Vo}\\
                & &  \hspace{9cm} n = 0,1,2,\cdots, \nonumber
\end{eqnarray}
where the variables with index 0 denotes classical fields
in the Lagrangian: 
$v_0 \equiv \phi$, $u_{0,\mu} \equiv \omega_{\mu}$, $b_0 \equiv B$,
$\xi_0\equiv\chi$, $\xi_{0,\mu}\equiv\chi_\mu$ 
and $\tilde{\xi}_0\equiv\tilde{\chi}$ 
and thus ${\cal{V}}_0 = {\cal{A}}$.
The variables with index 1 are the original gauge parameters as in 
eqs.(\ref{eqn:dchi}) and those with $n\ (>1)$ 
become the $n$-th reducibility parameters.
The minus signs in eq.(\ref{eqn:Vo}) are chosen for the later convenience. 
Then the transformation of ${\cal{V}}_n$
\begin{equation}
  \delta {\cal{V}}_n  = 
             (-)^n [ \ Q + {\cal{A}} \ , \ {\cal{V}}_{n+1} \ ]_{(-)^{n+1}},
                            \qquad  n = 0,1,2,\cdots, 
\label{eqn:ir}
\end{equation}
satisfies the on-shell relation
\begin{eqnarray}
  \delta(\delta{\cal{V}}_n) & = &
  \delta {\cal{V}}_n\Bigr\vert_{{\cal V}_{n+1}\rightarrow 
                                   {\cal V}_{n+1}+\delta{\cal V}_{n+1}} 
                                   - \delta {\cal{V}}_n \nonumber \\
        & = & (-)^n [ \ Q + {\cal{A}} \ , \ 
                     \delta {\cal{V}}_{n+1} \ ]_{(-)^{n+1}} \nonumber \\
        & = & (-)^n \left[ \ Q + {\cal{A}} \ , \ 
                 (-)^{n+1} [ \ Q + {\cal{A}} \ , \ {\cal{V}}_{n+2} \ 
                     ]_{(-)^{n+2}} \ \right]_{(-)^{n+1}}  \nonumber \\
        & = &  - [ \ {\cal{F}} \ , \ {\cal{V}}_{n+2} \ ]  \nonumber  \\
        & = & 0,                     
\label{eqn:irr}
\end{eqnarray}
where we have used the equation of motion (\ref{eqn:f-curv}).
In the above equations, $[ \ , \ ]_{(-)^n}$ is a commutator for odd $n$ 
and an anticommutator for even $n$.
Since the transformation (\ref{eqn:ir}) for $n=0$ represents the gauge 
transformation, eq.(\ref{eqn:irr}) implies that the gauge transformation
is infinitely reducible. We give explicit reducibility transformations
in the component form for the later use:
\begin{equation}
 \begin{array}{rcl}
 \delta v_n &=& [\phi,v_{n+1}]_{(-)^{n+1}} 
                +(-)^{n+1} \{\chi,\xi_{n+1}\}, \\
 \delta u_{n \mu} &=& \partial_\mu v_{n+1} + [ \omega_\mu , v_{n+1} ]
                     - [ \phi , u_{n+1 \mu} ]_{(-)^n} \\
                  & &+(-)^{n+1}[\chi_\mu,\xi_{n+1}]_{(-)^{n+1}}
                     +(-)^{n+1}\{\chi,\xi_{n+1\mu}\},  \\
 \delta b_n &=& \epsilon^{\mu \nu}( \partial_\mu u_{n+1 \nu} + 
                                   [ \omega_\mu , u_{n+1 \nu} ] ) 
                + [ B,v_{n+1}]_{(-)^{n+1}} + [ \phi,b_{n+1} ]_{(-)^{n+1}} \\
            & & +(-)^n \epsilon^{\mu \nu}[\chi_\mu,\xi_{n+1\nu}]_{(-)^n}
                +(-)^{n+1}\{\tilde{\chi},\xi_{n+1}\}
                +(-)^{n+1}\{\chi,\tilde{\xi}_{n+1}\}, \\
 \delta \xi_n &=& [\phi,\xi_{n+1}]_{(-)^n} 
                  +(-)^n [\chi,v_{n+1}], \\
 \delta \xi_{n\mu} &=& \partial_\mu \xi_{n+1} + [ \omega_\mu , \xi_{n+1} ]
                       -[\phi,\xi_{n+1\mu}]_{(-)^{n+1}} \\
                   & & +(-)^n [\chi_\mu, v_{n+1}]_{(-)^{n+1}}
                       +(-)^n[\chi,u_{n+1\mu}],  \nonumber \\ 
 \delta \tilde{\xi}_n &=& \epsilon^{\mu \nu} ( \partial_\mu \xi_{n+1 \nu} + 
                                              [ \omega_\mu , \xi_{n+1 \nu} ] ) 
                          + [B,\xi_{n+1}]_{(-)^n} 
                          + [\phi,\tilde{\xi}_{n+1}]_{(-)^n} \\
                 & & +(-)^{n+1}\epsilon^{\mu \nu}[\chi_\mu,u_{n+1\nu}]_{(-)^n}
                     +(-)^n [\tilde{\chi},v_{n+1}]
                     +(-)^n [\chi,b_{n+1}], \\
                 &&   \hspace*{9.7cm} n = 1, 2, 3, \cdots. 
\label{eqn:rxi} 
 \end{array} 
\end{equation}
It is also important to recognize that $v_n$, $u_{n\mu}$, $b_n$ are 
bosonic parameters while 
$\xi_n$, $\xi_{n,\mu}$, $\tilde{\xi_n}$ are fermionic parameters.

Actually the infinite on-shell reducibility is a common feature 
of generalized Chern-Simons theories with nonabelian gauge algebras 
in arbitrary dimensions, which can be understood 
by the fact that (\ref{eqn:irr}) is the relation 
among the generalized gauge fields and parameters.
Thus generalized Chern-Simons theories add another category 
of infinitely reducible
systems to known examples like Brink-Schwarz superparticle~\cite{bs},
Green-Schwarz superstring~\cite{gs}
and covariant string field theories~\cite{w2}.
It should be noted that this theory is infinitely reducible though
it contains only a {\it finite} number of fields of {\it finite} rank 
antisymmetric tensors.
Brink-Schwarz superparticle and Green-Schwarz superstring are 
the similar examples in the sense that they contain only a finite number 
of fields yet are infinitely reducible.
In the present case the infinite reducibility is understood 
from the following facts;
firstly, the highest form degrees of ${\cal{V}}_n$ is unchanged from that of
${\cal{V}}_{n-1}$ in eq.(\ref{eqn:ir}) 
since the generalized gauge field $\cal{A}$ contains 
the zero form gauge field $\phi$ and $\chi$,
secondly, the generalized Chern-Simons actions possess the same functional 
form (\ref{eqn:gCSa}) as the ordinary Chern-Simons action 
and thus have the vanishing curvature condition as the equation of motion;
${\cal{F}} = 0$ (\ref{eqn:f-curv}).
Thus the equations in (\ref{eqn:ir}) representing the infinite reducibility 
have the same form at any stage $n$, 
except for the difference between commutators and anticommutators. 
Algebraically, the structure of infinite reducibility resembles 
that of string field theories of a Chern-Simons form.

Before closing this section, 
we compare the generalized Chern-Simons theory  of the abelian 
$gl(1,{\bf R})$ algebra with the model of nonabelian algebra.
In the abelian case commutators in the gauge transformations vanish 
while anticommutators remain. 
Furthermore the field $\tilde{\chi}$
disappears from the classical Lagrangian.
Then we can consistently put all transformation parameters to be zero 
except for $v_1$, $u_{1 \mu}$, $v_2$ and $\xi_1$.
Thus the abelian model can be treated as a first stage reducible system.
In particular the purely bosonic abelian model 
was explicitly  quantized as a first stage reducible system
in the previous paper~\cite{kos}.
In nonabelian cases, however, infinite reducibility is a universal 
and inevitable feature of the generalized Chern-Simons theories.

%%%%%%%%%%%%%%%%%%%%
\subsection{Minimal sector} 

In this section we present a construction of the minimal part 
of quantized action based on the Lagrangian formulation 
given by Batalin and Vilkovisky~\cite{bv}.

In the construction of Batalin and Vilkovisky, ghosts, ghosts for ghosts 
and the corresponding antifields are introduced according to the reducibility 
of the theory.
We denote a minimal set of fields by $\Phi^A$ which include classical fields 
and ghost fields, and the corresponding  antifields by $\Phi_A^{\ast}$.
If a field has  ghost number $n$, its antifield has ghost number $-n-1$.
Then a minimal action is obtained by solving the master equation
\begin{eqnarray}
  (S_{min}(\Phi , \Phi^*),S_{min}(\Phi , \Phi^*)) &=& 0,
\label{eqn:me} \\
  (X,Y)&=&{\partial_r X \over \partial \Phi^A}
          {\partial_l Y \over \partial \Phi_A^*}
          -{\partial_r X \over \partial \Phi_A^*}
          {\partial_l Y \over \partial \Phi^A},
\label{eqn:ab}
\end{eqnarray}
with the boundary conditions
\begin{eqnarray}
  S_{min} \Bigr\vert_{\Phi^{\ast}_A = 0} & = & S_0, 
\label{eqn:bbc} \\
  \frac{\partial S_{min}}{\partial \Phi^{\ast}_{a_n}} 
                             \Bigr\vert_{\Phi^{\ast}_A = 0} 
           & = & Z^{a_n}_{a_{n+1}} \Phi^{a_{n+1}}, 
\label{eqn:bbd} \hspace{2cm} n = 0,1,2,\cdots, 
\end{eqnarray}
where $S_0$ is the classical action and $Z^{a_n}_{a_{n+1}} \Phi^{a_{n+1}}$ 
represents the $n$-th reducibility transformation where the reducibility 
parameters are replaced by the corresponding ghost fields.
In this notation, the relation with $n = 0$ in eq.(\ref{eqn:bbd}) 
corresponds to the gauge transformation. 
The BRST transformations of $\Phi^A$ and $\Phi_A^{\ast}$ 
are given by 
\begin{equation}
 s \Phi^A = (\Phi^A , S_{min}( \Phi , \Phi^{\ast} )), \hspace{1cm}
 s \Phi_A^{\ast} = (\Phi_A^{\ast} , S_{min}( \Phi , \Phi^{\ast} )). 
\label{eqn:br}
\end{equation}
Eqs.(\ref{eqn:me}) and (\ref{eqn:br}) assure that the BRST transformation 
is nilpotent and the minimal action is invariant under the transformation.
In usual cases, the master equation 
is solved order by order with respect to the ghost number.
Instead of solving an infinite set of equations due to the infinite
reducibility in the present case,
we can obtain the solution of the master equation (\ref{eqn:me}) by using
the characteristics of generalized Chern-Simons theory in which fermionic 
and bosonic fields, and odd and even forms, can be treated 
in a unified manner~\cite{kstu1}. 

First we introduce infinite fields
\begin{equation}
 \begin{array}{rcl}
 C_n^B, \ C_{n \mu}^B, \ \widetilde{C}_n^B = 
      \displaystyle{\frac{1}{2}} \epsilon^{\mu \nu} C_{n \mu \nu}^B, 
   &{\qquad}&  \\
 C_n^F, \ C_{n \mu}^F, \ \widetilde{C}_n^F = 
      \displaystyle{\frac{1}{2}} \epsilon^{\mu \nu} C_{n \mu \nu}^F, 
   &{\qquad}& n = 0, \pm1, \pm2,\cdots,\pm \infty,          
\raisebox{8mm}{} \label{eqn:cnc}
 \end{array}
\end{equation}
where the fields with index $B$ and $F$ are bosonic and fermionic,
respectively. The index $n$ indicates the ghost number of the field.  
The fields with ghost number $0$ are the classical fields
\begin{eqnarray*}
& & C_0^B = \phi, \ \ C_{0 \mu}^B = \omega_{\mu}, \ \ \widetilde{C}_0^B = B, 
            \nonumber \\
& & C_0^F = \chi, \ \ C_{0 \mu}^F = \chi_{\mu}, \ \ 
\widetilde{C}_0^F = \tilde{\chi}. 
\end{eqnarray*}
It is seen from eqs.(\ref{eqn:dchi}) and  
(\ref{eqn:rxi}) that fields content for ghosts and 
ghosts for ghosts in the minimal set is completed in the sector for $n>0$ 
while the necessary degrees of freedom for antifields are saturated for $n<0$. 
We will later identify fields with negative ghost numbers as antifields.
We now redefine a generalized gauge field $\widetilde{{\cal A}}$ 
in such a form of (\ref{eqn:ggf}) as it contains these infinite fields 
according to their Grassmann parity and form degrees:
\begin{eqnarray}
\widetilde{{\cal A}} &=& {\mbox{\bf 1}}\psi + {\mbox{\bf i}} \hat{\psi} +  
            {\mbox{\bf j}} A + {\mbox{\bf k}} \hat{A} 
            \ \in \Lambda_-,                           
\label{eqn:2d-ggf} \\
 &&\psi =  \sum_{n = -\infty}^{\infty} C_{n \mu}^F dx^{\mu},    \nonumber \\ 
 &&\hat{\psi}  =  \sum_{n = -\infty}^{\infty}  \left( C_n^F + 
             \frac{1}{2} C_{n \mu \nu}^F dx^{\mu} \wedge dx^{\nu} \right),
                                                                \nonumber \\
 &&A =  \sum_{n = -\infty}^{\infty} C_{n \mu}^B dx^{\mu},       \nonumber \\
 &&\hat{A} =  \sum_{n = -\infty}^{\infty}  \left( 
          C_n^B + \frac{1}{2} C_{n \mu \nu}^B dx^{\mu} \wedge dx^{\nu} \right).
                                                                \nonumber    
\end{eqnarray}
We then introduce a generalized action for $\widetilde{\cal A}$ as
\begin{eqnarray}
  \widetilde{S} & = &  \int  \ {\mbox{Tr}}^0_{\mbox{\bf {k}}} 
             \left( \frac {1}{2} \widetilde{{\cal A}}Q\widetilde{{\cal A}} 
                + \frac {1}{3} \widetilde{{\cal A}}^3 \right) 
\label{eqn:MA} \\
  & = & - \int {d^2}x {\mbox{Tr}}^0  
       \sum_{n=-\infty}^{\infty} 
       \left\{C_n^B \left(  \epsilon^{\mu \nu} \partial_{\mu}
         C_{-n \nu}^B \raisebox{6mm}{} \right. \right.  \nonumber \\
  & & \hspace{1cm}+\hspace{-3mm} \left.\left.
          \sum_{m=-\infty}^{\infty} \left( \epsilon^{\mu \nu}
        C_{m \mu}^B C_{-(m+n) \nu}^B + \{C_m^F,\widetilde{C}_{-(m+n)}^F \}
   - \epsilon^{\mu \nu} C_{m \mu}^F C_{-(m+n) \nu}^F 
                  \right) \right) \right. \nonumber \\
  & & \hspace{1cm}+\hspace{-3mm} \left.   
                \sum_{m=-\infty}^{\infty}
                       \widetilde{C}_{n}^B \left( C_m^F C_{-(m+n)}^F 
                    + C_m^B C_{-(m+n)}^B \right) \right.\nonumber \\
  & & \hspace{1cm}- C_{n}^F  \epsilon^{\mu \nu}
              \Bigl( \partial_{\mu} C_{-n\nu}^F + \sum_{m=-\infty}^{\infty}
                        [ C_{m \mu}^B , C_{-(m+n)\nu}^F ] \Bigr)
                                                    \Bigg\},
\label{eqn:Stilde}
\end{eqnarray}
where the upper index 0 on ${\mbox{Tr}}$ indicates to pick 
up only the part with ghost number 0.

One of the great advantage of generalized Chern-Simons formulation is 
that the quaternion valued gauge field and parameter which include different 
degrees of forms can be treated as if they were single gauge field and 
parameter.
Here we would like to identify the generalized action $\widetilde{S}$ 
with the minimal action itself. 
In order to obtain the similar algebraic structure as (\ref{eqn:br}) 
for the quaternion valued generalized gauge field, 
we heuristically introduce the following generalized antibracket:
\begin{equation}
  (X,Y)_{\lambda,{\bf k}}=\frac{1}{2}\int{\mbox{Tr}}_{\bf k}\bigg(
   {\delta_r X \over \delta {\cal A}}
          {\delta_l Y \over \delta {\cal A}_\lambda^*}
          -{\delta_r X \over \delta {\cal A}_\lambda^*}
          {\delta_l Y \over \delta {\cal A}} \bigg),
\end{equation}
where we define left-, right-functional derivative 
of the ``antifield'' ${\cal A}_\lambda^*$ by
\begin{eqnarray}
  {\delta_l X \over \delta {\cal A}_\lambda^*}
  \equiv {\delta_l ({\bf i}\lambda X) \over \delta {\cal A}}
  &=& -{\bf i}\lambda {\delta_l X \over \delta {\cal A}},\\
  {\delta_r X \over \delta {\cal A}_\lambda^*}
  \equiv {\delta_r ({\bf i}\lambda X) \over \delta {\cal A}}
  &=& {\bf i}\lambda {\delta_r X \over \delta {\cal A}},
\end{eqnarray}
where $\lambda$ is a fermionic scalar parameter with ghost number $-1$
and thus the relation, $\{ {\cal A},{\bf i}\lambda \}=0$ with 
${\cal A},{\bf i}\lambda \in \Lambda_-$, should be understood.
The role of ${\bf i}\lambda$ in the generalized antibracket 
could be understood as an analogy from the opposite Grassmann 
parity nature of antifields in the standard Batalin and 
Vilkovisky formulation.
 
In the following we need to use the generalized antibracket only for 
the two cases; 
i) both $X$ and $Y$ are functionals 
\Big($X={\displaystyle \int} {\mbox{Tr}}_{\bf k} f({\cal A}), \ f({\cal A}) 
\in \Lambda_-$\Big), 
ii) $X$ is a function ($X=f({\cal A})$) and $Y$ is a functional. 
In these cases it suffices to define the generalized functional 
derivative which satisfies the following two properties:
\begin{eqnarray}
1)&& \quad {\delta_{l,r} X \over \delta {\cal A}}
={\partial_{l,r} f({\cal A}) \over \partial {\cal A}} \quad 
{\hbox{for}} 
\quad X=\int \hbox{Tr}_{\bf k} f({\cal A}), 
 \quad (f({\cal A}) \in \Lambda_-),  
\label{eqn:fd1}\\
2)&& \quad \int \hbox{Tr}_{\bf k}\bigg( {\delta_{l,r} f({\cal A}) 
\over \delta {\cal A}} g({\cal A})\bigg) 
={\partial_{l,r} f({\cal A}) \over \partial {\cal A}}g({\cal A}),
\qquad  (f({\cal A}),g({\cal A}) \in \Lambda_-). 
\label{eqn:fd2}
\end{eqnarray}
In particular eq.(\ref{eqn:fd2}) implies  
$$
\int \hbox{Tr}_{\bf k}\bigg( {\delta_{l,r} {\cal A} \over \delta {\cal A}} 
g({\cal A})\bigg) 
=g({\cal A}) \quad {\hbox{for}} \quad g({\cal A}) \in \Lambda_- .
$$

By using the above properties of the generalized antibracket, we can show 
that the generalized action given in eq.(\ref{eqn:MA})
 is invariant under the following transformation 
 which is reminiscent of BRST transformation (\ref{eqn:br}), 
\begin{equation}
 \delta_{\lambda} \widetilde{{\cal A}} 
 \equiv (\widetilde{{\cal A}},\widetilde{S})_{\lambda,{\bf k}}
 = -\widetilde{{\cal F}} \ {\mbox{\bf i}} \lambda,
\label{eqn:dl}
\end{equation}
where $\widetilde{{\cal F}}$ is the generalized curvature (\ref{eqn:f-curv}) 
constructed out of $\widetilde{{\cal A}}$ and the fermionic 
parameter $\lambda$ with ghost number $-1$.
It should be understood that the same ghost number sectors must be equated 
in eq.(\ref{eqn:dl}).
Since $\widetilde{{\cal F}}$ and ${\bf i}\lambda$ belong to $\Lambda_+$
and $\Lambda_-$, respectively, their product in the right hand side 
of eq.(\ref{eqn:dl}) belongs to the same $\Lambda_-$ class as 
$\widetilde{\cal A}$.
The invariance of the action $\widetilde{S}$ under the transformation 
(\ref{eqn:dl}) can be checked by the manipulation 
\begin{eqnarray*}
 \delta_{\lambda} \widetilde{S} 
 & = & (\widetilde{S},\widetilde{S})_{\lambda,{\bf k}} \\
 & = & - \int \ {\mbox{Tr}}^0_{\mbox{\bf {k}}} 
        \left\{ ( Q \widetilde{{\cal A}} + \widetilde{{\cal A}}^2 ) 
                  \widetilde{{\cal F}} \ {\mbox{\bf {i}}} \lambda \right\} \\
 & = &  \int \ {\mbox{Tr}}^0_{\mbox{\bf {j}}} 
               ( \widetilde{{\cal F}} \widetilde{{\cal F}} ) \cdot \lambda \\
 & = &  \int \ {\mbox{Tr}}^0_{\mbox{\bf {j}}}
                   \left\{ Q ( \widetilde{{\cal{A}}} Q \widetilde{{\cal{A}}} 
              + \frac{2}{3} \widetilde{{\cal{A}}}^3 ) \right\}\cdot\lambda \\ 
& = & 0, 
\end{eqnarray*}
where the subscript ${\bf j}$ plays the similar role as the subscript 
${\bf k}$, {\it i.e.}, to pick up only the coefficient of ${\bf j}$ 
in the trace.
The change of the subscript ${\bf k}$ to ${\bf j}$ is necessary to take 
${\bf i}$ into account in the trace in accordance with 
${\bf j} {\bf i} = - {\bf k}$.
Here we have simply ignored the boundary term and thus the invariance 
is valid up to the surface term.

We now show that a right variation $s$ defined by 
$\delta_\lambda \widetilde{{\cal A}}=s\widetilde{{\cal A}} \lambda$
is the BRST transformation.
First of all this transformation is nilpotent:
\begin{eqnarray}
 s^2\widetilde{{\cal A}}\lambda_2\lambda_1
 \equiv \delta_{\lambda_2} \delta_{\lambda_1} \widetilde{{\cal A}} =  
        - \delta_{\lambda_2} \widetilde{{\cal F}} \ {\mbox{\bf i}} \lambda_1 = 
            - [ \ Q + \widetilde{{\cal A}} \ , \ \widetilde{{\cal F}} \ ] 
                               \lambda_2 \lambda_1 = 0,
\label{eqn:nilp}
\end{eqnarray}
where the generalized Bianchi identity is used
$$
 [ \ Q + \widetilde{{\cal A}} \ , \ \widetilde{{\cal F}} \ ] = 
    [ \ Q + \widetilde{{\cal A}} \ , \ ( \,Q + \widetilde{{\cal A}} \,)^2 \ ]
        = 0.
$$
Next we need to show that the transformation $s$ is realized as 
the antibracket form of (\ref{eqn:br}).
The invariance of $\widetilde{S}$ under (\ref{eqn:dl}) implies that 
$\widetilde{S}$ is indeed the minimal action if we make a proper 
identification of fields of negative ghost numbers with antifields.
It is straightforward to see that the BRST transformations 
(\ref{eqn:br}), both for fields and antifields, are realized 
under the following identifications with 
$S_{min}=\widetilde{S}$:
\begin{equation}
 \begin{array}{rclcrclc}
 C_{-n \mu}^F &=& \epsilon_{\mu \nu}^{-1} C_{n-1}^{B\nu \ast}, &\quad&
 C_{-n \mu}^B &=& \epsilon_{\mu \nu}^{-1} C_{n-1}^{F\nu \ast}, &\cropen
 C_{-n}^F     &=& {\widetilde{C}_{n-1}}^{B\ast},  &\quad&
 C_{-n}^B     &=& -{\widetilde{C}_{n-1}}^{F\ast}, &\cropen
 \widetilde{C}_{-n}^F &=& C_{n-1}^{B\ast}, &\quad&
 \widetilde{C}_{-n}^B &=& -C_{n-1}^{F\ast}, &\qquad n = 1,2,3,\cdots,
\label{eqn:idantif}
 \end{array}
\end{equation}
where $\epsilon_{\mu \nu}^{-1}$ is the inverse of $\epsilon^{\mu \nu}$, 
$\epsilon^{\mu \rho} \epsilon_{\rho \nu}^{-1} = \delta_{\nu}^{\mu}$\rlap.
\footnote{
%%%%%%%%%%footnote%%%%%%%%%%%%%%
To be precise the antifields are defined as \ 
 $C_n^{B\ast} = C^{B\ast}_{na} (\eta^{-1})^{ab} T_b,\cdots,$
with ${\mbox{Tr}} T_{a} T_{b} = \eta_{ab}$. 
%%%%%%%%%%%%%%%%%%%%%%%%%%%%%%%%
}
This shows that we have obtained a solution for the master equation 
(\ref{eqn:me}):
\begin{equation}
  \delta_{\lambda} S_{min} 
    = (S_{min},S_{min})_{\lambda,{\bf k}}
    = ( S_{min} , S_{min} ) \cdot \lambda = 0,
\label{eqn:master}
\end{equation}
where (~,~) is the original antibracket defined by (\ref{eqn:ab}).

It is easy to see that this solution satisfies 
the boundary conditions (\ref{eqn:bbc}) and (\ref{eqn:bbd}),
by comparing the gauge transformations 
(\ref{eqn:dchi}) and the reducibilities (\ref{eqn:rxi}) 
with the following expansion of $S_{min}$:
\begin{eqnarray*}
  S_{min} & = &  S_0+\int {d^2}x {\mbox{Tr}} \biggl\{
           \sum_{n=0}^{\infty} \Bigl\{ \ 
               C_{n}^{B\ast} \Bigl( [ \phi , C_{n+1}^F ]
                  -[\chi,C_{n+1}^B]\Bigr)
              + C_{n}^{F\ast}\Bigl( \{ \phi , C_{n+1}^B \}
                  +\{\chi,C_{n+1}^F\}\Bigr)                  \\
     & & \hspace{1cm} + 
         C_{n}^{B\mu \ast} \left( \ \partial_{\mu} C_{n+1}^F 
           + [ \omega_{\mu} , C_{n+1}^F ] 
           - \{ \phi , C_{n+1 \mu}^F\}
           - \{ \chi_{\mu} , C_{n+1}^B \} 
           - [ \chi , C_{n+1 \mu}^B] \ \right)                \\
     & & \hspace{1cm} + 
         C_{n}^{F\mu \ast} \left( \ \partial_{\mu} C_{n+1}^B 
           + [ \omega_{\mu} , C_{n+1}^B ] 
           - [ \phi , C_{n+1 \mu}^B]
           + \{ \chi_{\mu} , C_{n+1}^F \} 
           + \{ \chi , C_{n+1 \mu}^F\} \ \right)                \\
     & & \hspace{1cm} + 
            \widetilde{C}_{n}^{B\ast} \Bigl( 
            \epsilon^{\mu \nu} ( \partial_{\mu} C_{n+1 \nu}^F 
            + [ \omega_{\mu},C_{n+1 \nu}^F] + [\chi_\mu,C_{n+1\nu}^B]) 
                                          \\
     & & \hspace{2.5cm} 
             + [ B , C_{n+1}^F ]+ [\phi ,\widetilde{C}_{n+1}^F ]
             - [ \tilde{\chi} , C_{n+1}^B ]- [\chi ,\widetilde{C}_{n+1}^B ]
                                              \   \Bigr) \\
     & & \hspace{1cm} + 
            \widetilde{C}_{n}^{F\ast} \Bigl( 
            \epsilon^{\mu \nu} ( \partial_{\mu} C_{n+1 \nu}^B 
            + [ \omega_{\mu},C_{n+1 \nu}^B] - [\chi_\mu,C_{n+1\nu}^F]) 
                                          \\
     & & \hspace{2.5cm} 
             + \{ B , C_{n+1}^B \}+ \{\phi ,\widetilde{C}_{n+1}^B \}
             + \{ \tilde{\chi},C_{n+1}^F \}+\{\chi ,\widetilde{C}_{n+1}^F \}
                                           \   \Bigr) \ \Bigr\} 
                               + \cdots\cdots  \biggr\}.
\end{eqnarray*}
Thus the action $S_{min}=\widetilde{S}$ 
with the identification (\ref{eqn:idantif}) is the correct solution 
of the master equation for the generalized Chern-Simons theory.
The signs in eq.(\ref{eqn:Vo}) have been chosen so that the boundary 
conditions are satisfied without additional signs in the definition
of ghost fields in eq.(\ref{eqn:2d-ggf}).

For completeness we give explicit forms of the BRST transformations 
of the minimal fields
\begin{eqnarray}
 s C_{n}^B & = & - \sum_{m = -\infty}^{\infty} 
                                 [ C_{m}^F , C_{n-m+1}^B ], 
                                               \nonumber \\ 
 s C_{n}^F & = &  \sum_{m = -\infty}^{\infty} 
                 \left( \frac{1}{2} \{ C_{m}^B , C_{n-m+1}^B \} + 
                        \frac{1}{2} \{ C_{m}^F , C_{n-m+1}^F \} \right),
                                                \nonumber \\
 s C_{n \mu}^B & = & \partial_{\mu} C_{n+1}^F + \sum_{m = -\infty}^{\infty}
                       \left( [ C_{m \mu}^B , C_{n-m+1}^F ] - 
                             \{ C_{m \mu}^F , C_{n-m+1}^B \} \right), 
                                                \nonumber \\
 s C_{n \mu}^F & = & \partial_{\mu} C_{n+1}^B + \sum_{m = -\infty}^{\infty}
                       \left( [ C_{m \mu}^B , C_{n-m+1}^B ] + 
                             \{ C_{m \mu}^F , C_{n-m+1}^F \} \right),
\label{eqn:brst} \\
 s \widetilde{C}_{n}^B & = & \epsilon^{\mu \nu} \partial_{\mu} C_{n+1 \nu}^F
              + \sum_{m = -\infty}^{\infty} \Big( \epsilon^{\mu \nu} 
                   [ C_{m \mu}^B,C_{n-m+1 \nu}^F] \nonumber \\ 
       & &    \hspace{4cm}- [ \widetilde{C}_{m}^F , C_{n-m+1}^B ] - 
                     [ C_{m}^F , \widetilde{C}_{n-m+1}^B ] \Big),
                                                  \nonumber  \\
 s \widetilde{C}_{n}^F & = & 
                 \epsilon^{\mu \nu} \partial_{\mu} C_{n+1 \nu}^B \nonumber 
               +\sum_{m = -\infty}^{\infty} \Bigg( 
          \frac{1}{2} \epsilon^{\mu \nu} [ C_{m \mu}^B , C_{n-m+1 \nu}^B ] 
                  - \frac{1}{2} \epsilon^{\mu \nu} 
                [ C_{m \mu}^F , C_{n-m+1 \nu}^F ]  \nonumber \\ 
       & &   \hspace{4cm} + \{ C_{m}^B , \widetilde{C}_{n-m+1}^B \} + 
              \{ C_{m}^F , \widetilde{C}_{n-m+1}^F \} \Bigg),  \nonumber 
\end{eqnarray}
where the identification (\ref{eqn:idantif}) should be understood.

It is critical in our construction of the minimal action that the action 
of the generalized theory possesses the same structure as the Chern-Simons 
action and fermionic and bosonic fields are treated in a unified manner. 
It is interesting to note that the starting classical action
and the minimal action 
which includes the infinite series of bosonic and fermionic fields 
have the same form of (\ref{eqn:gCSa}) with the replacement 
${\cal{A}} \rightarrow \widetilde{\cal{A}} $.  
This is reminiscent of string field theories 
whose actions have the Chern-Simons form:
a string field contains infinite series of ghost fields and antifields.
The minimal action also takes the same Chern-Simons form~\cite{w2}.
It is also worth mentioning that there are other examples where
classical fields and ghost fields are treated in a unified way~\cite{ba}.

It is obvious that the minimal action for generalized Chern-Simons theory 
in arbitrary even dimensions can be constructed in the same way 
as in the two-dimensional case  
because the classical action (\ref{eqn:gCSa}), gauge symmetries 
(\ref{eqn:ggt}), 
reducibility transformations (\ref{eqn:ir}), 
the minimal action (\ref{eqn:MA}) and BRST transformations 
$ s\widetilde{\cal{A}} = - \widetilde{\cal{F}} {\mbox{\bf{i}}} $ 
are described by using generalized fields and parameters.

%%%%%%%%%%%%%%%%%%%%%
\subsection{Gauge-fixed action} 

The gauge degrees of freedom are fixed by introducing a nonminimal action 
which must be added to the minimal one, and choosing a suitable gauge fermion.
Though the number of gauge-fixing conditions is determined in accordance with 
the ``real'' gauge degrees of freedom, 
we can prepare a redundant set of gauge-fixing conditions and then
compensate the redundancy by introducing extraghosts.
Indeed Batalin and Vilkovisky gave a general prescription to construct
a nonminimal sector by this procedure~\cite{bv}.
This prescription is, however,
inconvenient in the present case since it leads to
a doubly infinite number of fields; antighosts, 
extraghosts,$\cdots$, where ``doubly infinite'' means the infinities 
both in the vertical direction and the horizontal direction 
in the triangular tableau of ghosts.      
In the case of the purely bosonic model,
we could find gauge-fixing conditions so that 
such extra infinite series do not appear while propagators for all fields 
be well-defined. 
This type of gauge-fixing prescription 
which is unconventional for the Batalin-Vilkovisky
 formulation is known, for example, 
in a quantization of topological Yang-Mills theory~\cite{lp,bbrt}.
Those gauge-fixing conditions are easily extended
to the present case and we can adopt 
the standard Landau type gauge-fixing for the vector and antisymmetric 
tensor fields in each sector of the ghost number,
which is sufficient to make a complete gauge-fixing.  

We introduce the nonminimal action 
\begin{eqnarray}
 S_{nonmin} &=& \int d^2 x \sum_{n=1}^{\infty} {\mbox{Tr}} 
      \Bigl( \bar{C}_{n}^{F\ast} b_{n-1}^B+\bar{C}_{n}^{B\ast} b_{n-1}^F
            + \bar{C}_{n \mu}^{F\ast} b_{n-1}^{B\mu}
            + \bar{C}_{n \mu}^{B\ast} b_{n-1}^{F\mu}\nonumber\\
 & & \hspace{2.5cm}
            + \eta_{n-1}^{F\ast} \pi_{n}^B
            + \eta_{n-1}^{B\ast} \pi_{n}^F \Bigr),
\label{eqn:2d-nonmin}
\end{eqnarray}
where the ghost number of nonminimal fields is $n$ for $\eta_n^{B,F}$, 
$\pi_n^{B,F}$ and $-n$ for $\bar{C}_n^{B,F}$, $\bar{C}^{B,F\mu}$, $b_n^{B,F}$, 
and the corresponding antifields possess ghost number $-n-1$ and $n-1$, 
respectively.
The indices $B$ and $F$ represent the Grassmannian property
of fields as before.
The BRST transformations of these fields are defined 
by this nonminimal action,
%
%%%%%
\renewcommand{\arraystretch}{1.3}
%%%%%
%
\begin{eqnarray}
 \begin{array}{rclcrclcrcl}
  s \bar{C}_n^{F,B} &=& b_{n-1}^{B,F}, &\quad&
  s b_{n-1}^{B,F}   &=& 0, & & & &  \cropen
  s \bar{C}_n^{F,B\mu} &=& b_{n-1}^{B,F\mu}, &\quad&
  s b_{n-1}^{B,F\mu} &=& 0, & & & &\cropen
  s \eta_{n-1}^{F,B} &=& \pi_{n}^{B,F}, &\quad&
  s \pi_{n}^{B,F} &=& 0, & & & &\cropen
  s \bar{C}_n^{F,B\ast} &=& 0, &\quad&
  s b_{n-1}^{B\ast} &=& -\bar{C}_n^{F\ast}, &\quad&
  s b_{n-1}^{F\ast} &=& \bar{C}_n^{B\ast}, \cropen
  s \bar{C}_{n \mu}^{F,B\ast} &=& 0, &\quad&
  s b_{n-1 \mu}^{B\ast} &=& -\bar{C}_{n \mu}^{F\ast}, &\quad&
  s b_{n-1 \mu}^{F\ast} &=& \bar{C}_{n \mu}^{B\ast}, \cropen
  s \eta_{n-1}^{F,B\ast} &=& 0, &\quad&
  s \pi_{n}^{B\ast} &=& -\eta_{n-1}^{F\ast}, &\quad&
  s \pi_{n}^{F\ast} &=& \eta_{n-1}^{B\ast}. 
\label{eqn:be} 
 \end{array}
\end{eqnarray}
%
%%%%%
\renewcommand{\arraystretch}{1.6}
%%%%%
%
The suffix $F$, $B$, which denotes fermionic or bosonic property of 
each ghost fields,
represents both relations with the given order. 
Next we adopt the following gauge fermion $\Psi$ 
which leads to a Landau type gauge-fixing,
\begin{eqnarray}
 \Psi &=& \int d^2 x \sum_{n=1}^{\infty} {\mbox{Tr}} \Bigl( 
          \bar{C}_{n}^B \partial^{\mu} C_{n-1 \mu}^F
          +\bar{C}_{n}^F \partial^{\mu} C_{n-1 \mu}^B
          + \bar{C}_{n}^{B\mu} \epsilon_{\mu \nu}^{-1} \partial^{\nu}
                      \widetilde{C}_{n-1}^F \nonumber\\
   & & \hspace{2cm}
          + \bar{C}_{n}^{F\mu} \epsilon_{\mu \nu}^{-1} \partial^{\nu}
                      \widetilde{C}_{n-1}^B
          + \bar{C}_{n}^{B\mu} \partial_{\mu} \eta_{n-1}^F 
          + \bar{C}_{n}^{F\mu} \partial_{\mu} \eta_{n-1}^B \Bigr), 
\label{eqn:ps}
\end{eqnarray}
where we assume a flat metric for simplicity. 
Then the antifields can be eliminated by equations 
$\Phi_A^* = \frac{\partial \Psi}{ \ \partial \Phi^A}$ 
\begin{eqnarray}
 \begin{array}{rclcrcl}
 C_n^{F,B\ast} &=& 0, & &
 C_n^{F,B\mu \ast} &=& -\partial^{\mu} \bar{C}_{n+1}^{B,F}, \\
 \widetilde{C}_n^{F,B\ast} &=& \epsilon_{\mu \nu}^{-1} \partial^{\mu} 
                               \bar{C}_{n+1}^{B,F\nu}, & &
 \bar{C}_{n+1}^{F,B\ast} &=& \partial^{\mu} C_{n \mu}^{B,F}, \\
 \bar{C}_{n+1 \mu}^{F,B\ast} &=& \epsilon_{\mu \nu}^{-1} \partial^{\nu} 
                                 \widetilde{C}_n^{B,F} 
                                 + \partial_{\mu} \eta_n^{B,F}, &\qquad &
 \eta_{n-1}^{F,B\ast} &=& -\partial_{\mu} \bar{C}_n^{B,F\mu}, \\
 & & & & & & \hspace*{2cm} n = 0,1,2,\cdots.   
\label{eqn:ioat} 
 \end{array}
\end{eqnarray}
The complete gauge-fixed action $S_{tot}$ is 
\begin{eqnarray}
 S_{tot} = S_{min}|_{\Sigma} + S_{nonmin}|_{\Sigma},  
\label{eqn:gfa}
\end{eqnarray}
where $\Sigma$ is a surface defined by 
eq.(\ref{eqn:ioat}).
This action is invariant under the on-shell nilpotent BRST transformations
(\ref{eqn:brst}) and (\ref{eqn:be}) 
in which the antifields are eliminated by substituting 
eqs.(\ref{eqn:ioat}).
It can be seen that the propagators of all fields are well-defined, 
by writing the kinetic terms and the gauge-fixing terms in  
\begin{eqnarray}
  S_{tot} & = & \int d^2 x {\mbox{Tr}} \Big\{ \ 
               - \phi \epsilon^{\mu \nu} \partial_{\mu} \omega_{\nu} 
               +  \partial^{\mu} \omega_{\mu} b_0^B + 
               \epsilon_{\mu \nu}^{-1} \partial^{\nu} B \ b_0^{B\mu}  
                                                          \nonumber  \\
         & &  \ \ \ \ \ \ \ \ \ \ \ \  \ \  
                + \chi \epsilon^{\mu \nu} \partial_{\mu} \chi_{\nu} 
               +  \partial^{\mu} \chi_{\mu} b_0^F + 
               \epsilon_{\mu \nu}^{-1} 
                      \partial^{\nu} \tilde{\chi} \ b_0^{F\mu}   
                                                          \nonumber \\
         & &  \ \ \ \ \ \ \ \ \ \ \ \  \ \  
                + \sum_{G=F,B}\sum_{n=1}^{\infty} \Big( 
                  - \partial^{\mu} \bar{C}_n^G \partial_{\mu} C_n^G
                  - \frac{1}{2} ( \partial^{\mu} \bar{C}_n^{G\nu} 
                                 - \partial^{\nu} \bar{C}_n^{G\mu} ) 
                                ( \partial_{\mu} C_{n \nu}^G 
                                 - \partial_{\nu} C_{n \mu}^G ) \Big) 
                                                           \nonumber  \\
        & & \ \ \ \ \ \ \ \ \ \ \ \ \ \ 
               +  \sum_{G=F,B}\sum_{n=1}^{\infty} \left( 
                \partial^{\mu} C_{n \mu}^G b_{n}^G + 
                \epsilon_{\mu \nu}^{-1} \partial^{\nu} 
                        \widetilde{C}_{n}^G b_n^{G\mu}
                      + \partial_{\mu} \eta_{n-1}^G  b_{n-1}^{G\mu} - 
               \partial_{\mu} \bar{C}_n^{G\mu} \pi_n^G \right) 
                                                           \nonumber \\
         & & \ \ \ \ \ \ \ \ \ \ \ \ \ \ 
                      + \ \mbox{ interaction \ terms } \ \Big\}.
\label{eqn:totkt}
\end{eqnarray}
Thus the gauge fermion (\ref{eqn:ps}) is a correct choice
and the gauge degrees of freedom are fixed completely.
We can consistently determine the hermiticity of the fields with a convention 
$\lambda^{\dagger} = - \lambda$ in eq.(\ref{eqn:dl})\rlap.\footnote{
%%%%%%footnote%%%%%%
Hermiticity conditions;
\begin{eqnarray*}
  & & C_n^{F,B\dagger}  =  - C_n^{F,B}, \ \ 
  C_{n \mu}^{F\dagger} = C_{n \mu}^F, \ \ 
  C_{n \mu}^{B\dagger} = -C_{n \mu}^B, \ \ 
  \widetilde{C}_n^{F,B\dagger} = \widetilde{C}_n^{F,B}, \\ 
  & & \bar{C}_n^{F\dagger} = \bar{C}_n^F, \ \ 
  \bar{C}_n^{B\dagger} = -\bar{C}_n^B, \ \ 
  \bar{C}_n^{F,B\mu \dagger} = - \bar{C}_n^{F,B\mu}, \\
  & & b_n^{F,B\dagger}  =  - b_n^{F,B}, \ \ 
   b_n^{F\mu \dagger} = - b_n^{F\mu}, \ \ 
   b_n^{B\mu \dagger} = - b_n^{B\mu}, \ \ 
  \eta_n^{F,B\dagger} = \eta_n^{F,B}, \ \ 
  \pi_n^{F\dagger} = \pi_n^F \ \
  \pi_n^{B\dagger} = -\pi_n^B .
\end{eqnarray*}
%%%%%%%%%%%%%%%%%%%%
}

Here comes a possible important comment.
There is a common feature for models of infinitely reducible systems.
When the number of reducibility parameters at each level is the 
same as that of gauge parameters, the number of the ``real'' 
gauge degrees of freedom 
is the half of the original degrees of freedom.
The known examples of this type, Brink-Schwarz superparticle and 
Green-Schwarz superstring, have this characteristics~\cite{bs,gs}. 
In the present two-dimensional model, there are eight parameters 
$v_{n}$, $u_{n \mu}$, $b_{n}$, $\xi_n$, $\xi_{n\mu}$ and $\tilde{\xi}_n$ 
for each stage of the reducibility.
The ``real'' number of gauge-fixing conditions is $6-2=4$,
where six gauge-fixing conditions 
${\partial}^{\mu} C_{n-1 \mu}^{F,B} = 0$,
$ {\epsilon}^{-1}_{\mu \nu} {\partial}^{\nu} {\widetilde{C}}_{n-1}^{F,B} = 0 $ 
are linearly dependent due to 
$ {\partial}^{\mu} ( {\epsilon}^{-1}_{\mu \nu} 
                    {\partial}^{\nu} {\widetilde{C}}_{n-1}^{F,B} ) = 0 $ 
and thus we needed to impose an extra condition 
$ {\partial}_{\mu} {\bar{C}}^{F,B\mu}_{n} = 0 $.

%%%%%%%%%%%%%%%%%%%%%
\section{Quantization in Hamiltonian formulation} 

\setcounter{equation}{0}
\setcounter{footnote}{0}

In this section we investigate the quantization of the same model 
(\ref{eqn:ba}) 
in the Hamiltonian formulation and show that the gauge-fixed action 
obtained from the Hamiltonian formulation coincides with that of the 
Lagrangian 
formulation if we make a proper choice of gauge fermion and a suitable 
identification of ghost fields. 
One of the important aim of carrying out the quantization of the same 
model in the Hamiltonian formulation is to see how the regularity violating 
constraints can be 
interpreted in the Hamiltonian formulation. 
In the previous paper on the analysis of the abelian version of the present 
model, it was pointed out that a physical degree of freedom which does not 
exist in the classical level appears in the quantum level and the origin 
of the appearance is essentially related to the violation of 
regularity~\cite{kos}. 
This situation is unchanged in the nonabelian version of the present 
model and the Hamiltonian formulation of the quantization confirms the 
result.

%%%%%%%%%
\subsection{Purely bosonic model}

For simplicity we first investigate the case where the classical action 
includes only bosonic fields.
Incorporation of fermionic classical fields can be done straightforwardly 
and will be explained in the next subsection. 
Following to the standard procedure, we obtain canonical momenta 
from the action (\ref{eqn:ba}) where fermionic classical fields are omitted, 
\begin{eqnarray}
  \pi_{\phi} & = & 0, \label{eqn:pip} \\
  \pi_{\omega_0} & = & 0, \label{eqn:pio0} \\
  \pi_{\omega_1} & = & -\phi, \label{eqn:pio1} \\
  \pi_B & = & 0. \label{eqn:pib}
\end{eqnarray}
All of these equations give primary constraints.
The canonical Hamiltonian following from the Lagrangian becomes 
$$
  H_C = \int dx^1 \mbox{Tr} \Big\{ - \phi D_1 \omega_0 + \phi^2 B \Big\},
$$
where $D_1 \equiv \partial_1 + [\omega_1 , \ ]$ is the space component 
of the covariant derivative.
This Hamiltonian and the above primary constraints with Lagrange multipliers 
define the total Hamiltonian
$$
  H_T = \int dx^1 \mbox{Tr} \left\{ - \phi D_1 \omega_0 + \phi^2 B 
            + \lambda_{\phi} \pi_{\phi}  + \lambda_{\omega_0} \pi_{\omega_0} 
            + \lambda_{\omega_1} ( \pi_{\omega_1} + \phi ) 
            + \lambda_B \pi_B \right\}.
$$
According to the ordinary Dirac's procedure, we have to check the consistency 
of the constraints.
The results are 
\begin{eqnarray}
  \partial_0 \pi_{\phi} & = & D_1 \omega_0 - \{ \phi , B \} 
                                   - \lambda_{\omega_1} = 0, 
\label{eqn:lo1} \\
  \partial_0 \pi_{\omega_0} & = & - D_1 \phi = 0,  \\
  \partial_0 ( \pi_{\omega_1} + \phi ) & = & 
             - [ \phi , \omega_0 ]  + \lambda_{\phi} = 0, 
\label{eqn:lp} \\
  \partial_0 \pi_B & = & - \phi^2 = 0.
\end{eqnarray}
Two Lagrange multipliers, $\lambda_{\omega_1}$ and $\lambda_{\phi}$, 
are determined from (\ref{eqn:lo1}) and (\ref{eqn:lp}).
After the substitution of these expressions into the total Hamiltonian,
we obtain
\begin{equation}
  H_T = \int dx^1 \mbox{Tr} \left\{
          - \omega_0 ( D_1 \pi_{\omega_1} - [ \pi_{\phi} , \phi ] ) 
          - ( \phi^2 + \{ \pi_{\omega_1} , \phi \} ) B 
          + \lambda_{\omega_0} \pi_{\omega_0} 
          + \lambda_B \pi_B \right\}.
\label{eqn:th}
\end{equation}
At the same time, we have found secondary constraints
\begin{eqnarray}
  D_1 \phi & = & 0, \label{eqn:sec1} \\
  - \phi^2 & = & 0. \label{eqn:sec2}
\end{eqnarray}
The consistency check of these constraints gives no further relations.
Thus we have obtained the set of the constraints 
eqs.(\ref{eqn:pip})$-$(\ref{eqn:pib}) and 
eqs.(\ref{eqn:sec1}) and (\ref{eqn:sec2}).

It should be noted that the constraints (\ref{eqn:sec1}) and (\ref{eqn:sec2}) 
violate Dirac's regularity condition.
Constraints are called regular if any function of canonical variables 
vanishing on the constraint surface can be written as their ``linear'' 
combination, where the coefficients of the combination could be dependent on 
canonical variables.
More precisely to say, constraints $\{\Phi_1,\cdots,\Phi_M\}$ are 
regular if independent constraints $\Phi_1,\cdots,\Phi_{M'} (M'\le M)$ 
can be taken as the first $M'$ coordinates of the $N$-dimensional phase 
space in the vicinity of the $(N-M')$-dimensional constraint surface 
and thus $d\Phi_1\wedge \cdots \wedge d\Phi_{M'} \ne 0$ on the surface. 
In the models with nonabelian gauge algebras there are two 
possibilities depending on the chosen gauge algebra.
Firstly, if we take Clifford algebras $c(k,0)$ or $c(0,k)$ where the 
metric of the 
algebra is positive definite or negative definite, then the equation 
$\phi^2 = 0$ is equivalent to $\phi = 0$ 
which should be the constraint surface determined by (\ref{eqn:sec1}) and 
(\ref{eqn:sec2}). 
Neither (\ref{eqn:sec1}) nor (\ref{eqn:sec2}) can be taken as 
the above constraint for $M'=1$ because eq. (\ref{eqn:sec1}) does not 
imply $\phi=0$ while $\phi^2 = 0$ is not regular since $d\phi^2 = 0$ at 
$\phi=0$.
In this case we can replace these two constraints 
by a single equivalent constraint  
$$
  \phi = 0.
$$
Then we can separate the constraints into the first class and the second class.
With some redefinitions we can obtain a set of constraints as
\begin{eqnarray*}
  \mbox{second class} \quad & & \pi_{\phi} = 0, \ \ \phi = 0, \\
  \mbox{first class}  \quad & & \pi_{\omega_0} = 0, \\
                      \quad & & \pi_{\omega_1} = 0, \\
                      \quad & & \pi_B = 0.
\end{eqnarray*}
These constraints imply that there exist no dynamical variables, 
which is expected from the topological nature of the generalized 
Chern-Simons action.
The quantization of this system is trivial for the flat base manifold 
since there is no dynamical degrees of freedom.
Finite degrees of freedom may appear depending on the choice of a nontrivial 
topology for the base manifold. 
By adopting gauge-fixing conditions $\omega_0 = \omega_1 = B = 0$, 
all variables and thus Hamiltonian $H_T$ itself vanishes identically.
Thus we can conclude this theory is completely empty.
However this treatment does not lead to the quantized Lagrangian which 
we have obtained in the Lagrangian formulation.

The other situation is the case where $\phi^2 = 0$ is not equivalent to 
$\phi = 0$.
The specific examples of such algebras are Clifford algebras 
$c(m,n) \ ( m \neq 0, n \neq 0 )$.
In particular we now consider $c(2,1)$ algebra generated by $\{ T_a \} = 
\{ 1, \sigma_1, \sigma_2, i \sigma_3 \}$ where $\sigma_k$'s are Pauli matrices.
Then $\phi$ is expanded into components as
$\phi = \phi^s 1 + \phi^1 \sigma_1 + \phi^2 \sigma_2 + \phi^3 i \sigma_3$, 
and thus $\phi^2 = 0$ leads to 
$$
  (\phi^s)^2 + (\phi^1)^2 + (\phi^2)^2 - (\phi^3)^2 = 0, 
  \qquad  \phi^s \phi^k = 0.
$$
These equations are equivalent to 
\begin{eqnarray}
  \phi^s & = & 0, \\
  (\phi^1)^2 + (\phi^2)^2 - (\phi^3)^2 & = & 0.            
\label{eqn:pp}
\end{eqnarray}
Eq.(\ref{eqn:pp}) implies that the constraint surface is not a 
single but a branched hypersurface in the phase space since  
$\phi^3 = \pm \sqrt{(\phi^1)^2 + (\phi^2)^2}$.
In the case of $c(k,0)$ and $c(0,k)$ single regular constraint 
$\phi=0$ can replace the constraints (\ref{eqn:sec1}) and (\ref{eqn:sec2})
while one of the branch of 
$\phi^3 = \pm \sqrt{(\phi^1)^2 + (\phi^2)^2}$ 
cannot be taken as a regular constraint in the present case since 
one of the branced surface is not enough to specify whole 
the constraint surface and furthermore they themselves are singular 
at $\phi=0$.
Therefore it seems rather natural in the generalized Chern-Simons theory 
to adopt a quantization method different from the usual one, {\it i.e.}, 
a quantization based on regularity violating constraints that follow directly 
from the Lagrangian.
In the following we perform the Hamiltonian BRST quantization 
{\it \`{a} la} Batalin, Fradkin and Vilkovisky by using the regularity 
violating constraints.
It is interesting that in this treatment of Hamiltonian formulation 
with a suitable choice of gauge-fixing conditions, 
we can show that the gauge-fixed action obtained from the regularity violating 
constraint is just the same as 
the result of the Lagrangian formulation.
Though the Lagrangian and Hamiltonian constructions
are formally equivalent in usual cases as shown 
in refs. \cite{bv} and \cite{bfv}, we can show the equivalence
of two formulations in the present model only 
if we adopt the regularity violating constraints 
in the Hamiltonian formulation.

We first rearrange the constraints (\ref{eqn:pip})$-$(\ref{eqn:pib}), 
(\ref{eqn:sec1}) and (\ref{eqn:sec2}) into (\ref{eqn:pip})$-$(\ref{eqn:pib}) 
and 
\begin{eqnarray}
   - D_1 \pi_{\omega_1} - [ \pi_{\phi} , \pi_{\omega_1} ] & = & 0, 
\label{eqn:con1} \\
    - \pi^2_{\omega_1} & = & 0,  
\label{eqn:con2}
\end{eqnarray}
so that first class constraints, (\ref{eqn:pio0}), (\ref{eqn:pib}),
(\ref{eqn:con1}) and (\ref{eqn:con2}), and second class constraints, 
(\ref{eqn:pip}) and (\ref{eqn:pio1}), are separated.
We can now carry on without the variables $\phi$ and $\pi_{\phi}$ 
because of the second class constraints, that is, 
we can replace all $\phi$ by $- \pi_{\omega_1}$ and set $\pi_{\phi}$ 
to be zero by using Dirac's brackets.
We further adopt gauge conditions $\omega_0 = B = 0$ 
for the first class constraints (\ref{eqn:pio0}) and (\ref{eqn:pib}), 
for simplicity.
Then we can also eliminate $\omega_0$, $\pi_{\omega_0}$, $B$ and $\pi_B$ 
from the system.
After these manipulations, we have two phase space variables $\omega_1$ and 
$\pi_{\omega_1}$ with the first class constraints 
\begin{eqnarray}
  G_1 & \equiv & - D_1 \pi_{\omega_1} = 0, 
\label{eqn:g1} \\
  H_1 & \equiv & - \pi^2_{\omega_1} = 0. 
\label{eqn:h1}
\end{eqnarray}
and the total Hamiltonian (\ref{eqn:th}) vanishes completely.

This system with the first class constraints 
(\ref{eqn:g1}) and (\ref{eqn:h1}) violating the regularity condition 
is infinitely reducible, as it has been in the Lagrangian formulation:
\begin{eqnarray}
  G_n & \equiv & (-)^{n-1} D_1 H_{n-1} + [ \pi_{\omega_1} , G_{n-1} ]_{(-)^n}
                                                                   \nonumber \\
      & = & (-)^{n-1} [ D_1 \pi_{\omega_1} , H_{n-2} ]_{(-)^n} 
                 + [ \pi^2_{\omega_1} , G_{n-2} ] \nonumber \\
      & = & 0,       
 \label{eqn:red1} \\
  H_n & \equiv & [ \pi_{\omega_1} , H_{n-1} ]_{(-)^{n-1}} \nonumber \\
      & = & [ \pi_{\omega_1}^2 , H_{n-2} ] \nonumber \\
      & = & 0, \hspace{5cm} n = 2,3,4,\cdots.    
\label{eqn:red2} 
\end{eqnarray}
where we have used the constraints (\ref{eqn:g1}) and (\ref{eqn:h1}). 
In the case for $n=2$ the relations are satisfied trivially.

Here we comment on the number of linearly independent constraints.
If there are first class constraints the number of the physical degrees of 
freedom is reduced by twice the number of the first class constraints, 
since the constraint itself kills one degree and the corresponding 
gauge symmetry induces another unphysical degree.
In the counting of degrees of freedom, the multiplication of 
the gauge degrees of freedom by the dimension of the gauge algebra 
should be understood and will be omitted from the discussions. 
In the present model we have two phase space variables $\omega_1$ and 
$\pi_{\omega_1}$.
These degrees of freedom should be cancelled 
by one first class constraint so that the theory is topological and thus 
has no degrees of freedom even after the quantization.
Since we have two first class constraints (\ref{eqn:g1}) and (\ref{eqn:h1}), 
there should be one relation between them, in other words 
(\ref{eqn:g1}) and (\ref{eqn:h1}) are linearly dependent.
In fact what happened in the present model is that two reducibility 
conditions appear at each level.
According to the above argument the number of linearly independent 
equations for 
(\ref{eqn:red1}) and (\ref{eqn:red2}) at each level should also be one.
In order to compensate the over cancelled degrees of freedom, 
two reducibility conditions appear again and then the process repeats infinite 
times.
This is how the infinite reducibility appears in the Hamiltonian formulation. 

In the case of a simple constrained system, we can quantize the system 
without unphysical degrees of freedom by solving the constraints.
In many cases, however, we lose manifest invariance under important global 
symmetries and/or the locality.
Furthermore in the present case the solutions of the constraints 
(\ref{eqn:g1}) and 
(\ref{eqn:h1}) are determined according to the gauge algebra, 
as explained above. 
We will show that the action obtained in the Hamiltonian formulation 
coincides with the result of the Lagrangian formulation which contains 
infinitely many unphysical degrees of freedom, {\it i.e.}, ghost fields.
We adopt the Hamiltonian formulation given by Batalin, Fradkin and 
Vilkovisky which accommodates the reducibility of the system.
In this formulation a phase space is extended so as to contain ghosts and 
ghost momenta.
Then a nilpotent BRST differential is constructed and a physical phase space 
is defined as its cohomology which is a set of gauge invariant functions 
on the constraint surface.
The role of the ghost momenta is to exclude functions vanishing 
on the constraint surface from the cohomology and gauge variant functions 
are removed from the cohomology because of the action of the BRST differential 
for the ghosts.

First we introduce infinite ghosts and ghost momenta
\begin{equation}
 \begin{array}{llllllll}
 \eta_{2n}^B, & & \eta_{2n}^{'B}, 
              & & P_{2n}^B, & & P_{2n}^{'B},& \\
 \eta_{2n-1}^F, & & \eta_{2n-1}^{'F}, 
                & & P_{2n-1}^F, & & P_{2n-1}^{'F}, 
                & \hspace{2cm} n = 1,2,3,\cdots,
\label{eqn:gh}
 \end{array}
\end{equation}
where $\eta_n^{B,F}$ and $P_n^{B,F}$ correspond with 
the constraint (\ref{eqn:g1}) and the reducibility (\ref{eqn:red1}) 
and $\eta_n^{'B,F}$ and $P_n^{'B,F}$ 
with (\ref{eqn:h1}) and (\ref{eqn:red2}).
The ghost numbers of the $\eta_n^{B,F}$ 
and $\eta_n^{'B,F}$ are $n$ 
and those of $P_n^{B,F}$ and $P_n^{'B,F}$ 
are $-n$.
The fields with index $B$ and $F$ are bosonic and fermionic, respectively.
Although the way for constructing the BRST differential in the Hamiltonian 
formulation is systematic, it is complicated in the present case due to 
the infinite reducibility.
After the step by step construction according to the systematic procedure, 
we have found an elegant way of presenting the result.
Thus we first give the result we obtained 
and then show that it is just 
what the usual procedure leads to.
The construction which we carried out is similar to that 
in the Lagrangian formulation.
First we define a generalized gauge field $\widetilde{\cal A}$ from 
the ghosts, $\eta_n^{B,F}$ and $\eta_n^{'B,F}$, 
and the ghost momenta, $P_n^{B,F}$ and $P_n^{'B,F}$, by
\begin{eqnarray}
  \widetilde{\cal A} & = & \mbox{\bf 1} \psi + \mbox{\bf i} \hat{\psi} 
                         + \mbox{\bf j} A   + \mbox{\bf k} \hat{A}, 
\label{eqn:ah} \\
     & &      \psi  = \sum_{n=-\infty}^{\infty} \Gamma_{2n-1}^F dx^1 
                           \equiv \sum_{n=1}^{\infty} \left( 
                      \eta_{2n-1}^{'F} + P_{2n-1}^F \right) dx^1, 
                                                             \nonumber \\
     & &      \hat{\psi}  = \sum_{n=-\infty}^{\infty} \Pi_{2n-1}^F 
                \equiv \sum_{n=1}^{\infty} \left( 
                      \eta_{2n-1}^F + P_{2n-1}^{'F} \right), \nonumber \\
     & &      A  = \sum_{n=-\infty}^{\infty} \Gamma_{2n}^B dx^1
                   \equiv \bigg\{ \omega_1 + \sum_{n=1}^{\infty} \left( 
                      \eta_{2n}^{'B} - P_{2n}^B \right) \bigg\}dx^1, 
                                                             \nonumber \\
     & &      \hat{A}  =  \sum_{n=-\infty}^{\infty} \Pi_{2n}^B 
                      \equiv \pi_{\omega_1} + \sum_{n=1}^{\infty} \left(
                          \eta_{2n}^B + P_{2n}^{'B} \right). \nonumber
\end{eqnarray}
We next introduce the BRST differential $s$ as
\begin{eqnarray}
  s \widetilde{\cal A} = - \widetilde{\cal F} \mbox{\bf i} 
             = - ( Q' \widetilde{\cal A} + \widetilde{\cal A}^2 ) \mbox{\bf i},
\label{eqn:sham}
\end{eqnarray}
where $Q' \equiv {\bf j}dx^1 \partial_1$ is one dimensional exterior 
derivative which does not include the time derivative.
It should be noted that the BRST differential has the same form as 
the Lagrangian counterpart.
The nilpotency of the BRST differential, $s^2 = 0$, can be shown 
in the same way as in the Lagrangian formulation.
The explicit actions of the BRST differential are 
\begin{equation}
 \begin{array}{rcl}
  s \Pi_{2n}^B & = & {\displaystyle{\sum_{m=-\infty}^{\infty}}} 
                        [ \Pi_{2m}^B , \Pi_{2(n-m)+1}^F ], \\ 
  s \Pi_{2n-1}^F & = & {\displaystyle{\sum_{m=-\infty}^{\infty}}} \left( 
                \displaystyle{\frac{1}{2}} 
                   \{ \Pi_{2m}^B , \Pi_{2(n-m)}^B \} 
               +\displaystyle{\frac{1}{2}} 
                   \{ \Pi_{2m-1}^F , \Pi_{2(n-m)+1}^F \} \right), \\
  s \Gamma_{2n}^B & = & \partial_1 \Pi_{2n+1}^F 
                      + {\displaystyle{\sum_{m=-\infty}^{\infty}}} 
                \left( 
                 [ \Gamma_{2m}^B , \Pi_{2(n-m)+1}^F ] - 
                 \{ \Pi_{2m}^B , \Gamma_{2(n-m)+1}^F \} \right), \\
  s \Gamma_{2n-1}^F & = & \partial_1 \Pi_{2n}^B 
                      + {\displaystyle{\sum_{m=-\infty}^{\infty}}} 
                  \left( [ \Gamma_{2m}^B , \Pi_{2(n-m)}^B ] + 
                  \{ \Gamma_{2m-1}^F , \Pi_{2(n-m)+1}^F \} 
                                                   \right). 
\label{eqn:BRST}
 \end{array}  
\end{equation}

We now show that $s$ coincides with the BRST differential which is obtained 
by the formulation of Batalin, Fradkin and Vilkovisky.
In order to obtain the BRST differential we follow to the systematic 
procedure developed by Henneaux et al.~\cite{ht}. 
The BRST differential can be decomposed into 
$\displaystyle s = \delta + D + \sum_{k \geq 1} \stackrel{(k)}{s}$.
The homology of the Koszul-Tate differential $\delta$ is a set of functions 
on the constraint surface. 
The extended longitudinal differential $D$ is considered on the homology 
of the Koszul-Tate differential and its cohomology is a set of gauge 
invariant functions.
Finally $\stackrel{(k)}{s}$ is determined so that $s$ is nilpotent.
Then it is guaranteed by the (co)homological perturbation theory 
that the cohomology of the BRST differential is a set of gauge invariant 
functions on the constraint surface.

We first define the antighost number as 
\begin{eqnarray*}
  \mbox{antigh} ( \delta ) & = & -1, \ \ \mbox{antigh} ( D )  =  0, \ \ 
  \mbox{antigh} ( \stackrel{(k)}{s} )  =  k, \\
  \mbox{antigh} ( \omega_1 ) & = & \mbox{antigh} ( \pi_{\omega_1} ) = 0, \\ 
  \mbox{antigh} ( P_{2n}^B ) & = & \mbox{antigh} ( P_{2n}^{'B} ) = 2n, \\ 
  \mbox{antigh} ( P_{2n-1}^F ) & = & \mbox{antigh} ( P_{2n-1}^{'F} ) = 2n-1.
\end{eqnarray*}
The actions of the Koszul-Tate differential $\delta$ for the ghost momenta 
are read from eqs.(\ref{eqn:BRST}) by comparing antighost 
numbers:
\begin{equation}
 \begin{array}{rcl}
  \delta P_{2n+1}^F & = & D_1 P_{2n}^{'B} + [ \pi_{\omega_1} , P_{2n}^B ] \\
                    & &  -{\displaystyle{\sum_{m=1}^{n-1}}}
                          [ P_{2m}^B , P_{2(n-m)}^{'B} ] 
                         +{\displaystyle{\sum_{m=0}^{n-1}}}
                           \{ P_{2m+1}^F , P_{2(n-m)-1}^{'F} \}, \\
  \delta P_{2n}^B  &=& - D_1 P_{2n-1}^{'F} + \{ \pi_{\omega_1} , P_{2n-1}^F \} 
                                                                          \\
                   & &  + {\displaystyle{\sum_{m=1}^{n-1}}}
                          [ P_{2m}^B , P_{2(n-m)-1}^{'F} ] 
                        + {\displaystyle{\sum_{m=1}^{n-1}}}
                           \{ P_{2m}^{'B} , P_{2(n-m)-1}^F \}, \\
  \delta P_{2n+1}^{'F} & = & \{ \pi_{\omega_1} , P_{2n}^{'B} \} 
                             + {\displaystyle{\frac{1}{2} \sum_{m=1}^{n-1}}} 
                              \{ P_{2m}^{'B} , P_{2(n-m)}^{'B} \} 
                             + {\displaystyle{\frac{1}{2} \sum_{m=0}^{n-1}}} 
                              \{ P_{2m+1}^{'F} , P_{2(n-m)-1}^{'F} \}, \\
  \delta P_{2n}^{'B} & = & [ \pi_{\omega_1} , P_{2n-1}^{'F} ] + 
                           {\displaystyle{\sum_{m=1}^{n-1}}}
                           [ P_{2m}^{'B} , P_{2(n-m)-1}^{'F} ],
\label{eqn:del} 
 \end{array}
\end{equation}
in particular 
\begin{eqnarray*}
  \delta P_1^F  &=&  D_1 \pi_{\omega_1}, \\
  \delta P_1^{'F} &=&  \pi_{\omega_1}^2. 
\end{eqnarray*}
It is understood that these coincide with the actions of the Koszul-Tate 
differential constructed from the constraints (\ref{eqn:g1}) and 
(\ref{eqn:h1}) and the reducibilities (\ref{eqn:red1}) and (\ref{eqn:red2}).

We next consider the extended longitudinal differential $D$.
In cases with reducibilities, it is expanded into 
$\displaystyle D = \Delta + d + \sum_{k \leq -1} \stackrel{(k)}{D}$ 
where $d$ is called the longitudinal differential and $\Delta$ the auxiliary 
differential and the auxiliary degree is assigned as
\begin{eqnarray*}
  \mbox{aux}( \Delta ) & = & 1, \ \ \mbox{aux}( d ) = 0, \ \ 
                           \mbox{aux}( \stackrel{(k)}{D} ) = k, \\
  \mbox{aux}( \omega_1, \ \pi_{\omega_1} ) & = & 0 = 
                       \mbox{aux}( \eta_1^F, \ \eta_1^{'F} ), \\ 
  \mbox{aux}( \eta_{2n}^{B}, \ \eta_{2n}^{'B} ) & = & 2n-1, \\
  \mbox{aux}( \eta_{2n+1}^{F}, \ \eta_{2n+1}^{'F} ) & = & 2n,
                                     \hspace{2cm} n = 1,2,3,\cdots.
\end{eqnarray*}
The operations of the longitudinal differential $d$ for 
$\omega_1$, $\pi_{\omega_1}$, 
$\eta_1^F$ and $\eta_1^{'F}$ 
are determined from the form of the gauge transformations 
and the nilpotency of $d$:
\begin{equation}
 \begin{array}{rcl}
  d \omega_1 & = & D_1 \eta_1^F - \{ \pi_{\omega_1} , \eta_1^{'F} \}, \\ 
  d \pi_{\omega_1} & = & [ \pi_{\omega_1} , \eta_1^F ],  \\
  d \eta_1^F & = & {\eta_1^F}^2, \\
  d \eta_1^{'F} & = & \{ \eta_1^{'F} , \eta_1^F \}. 
\label{eqn:ld}
 \end{array} 
\end{equation}
The actions of $\Delta$ are obtained from the reducibilities (\ref{eqn:red1}) 
and (\ref{eqn:red2}) as
\begin{equation}
 \begin{array}{rcl}
  \Delta \omega_1 & = & 0 = \Delta \pi_{\omega_1}, \\
  \Delta \eta_{2n}^B & = & [ \pi_{\omega_1} , \eta_{2n+1}^F ],  \\ 
  \Delta \eta_{2n-1}^F & = & \{ \pi_{\omega_1} , \eta_{2n}^B \}, \\
  \Delta \eta_{2n}^{'B} & = & D_1 \eta_{2n+1}^F 
               - \{ \pi_{\omega_1} , \eta_{2n+1}^{'F} \},  \\
  \Delta \eta_{2n-1}^{'F} & = & D_1 \eta_{2n}^B 
               - [ \pi_{\omega_1} , \eta_{2n}^{'B} ]. 
\label{eqn:ad}
 \end{array}
\end{equation}
The transformation coefficients of the auxiliary differential $\Delta$ 
with respect to ghosts correspond to the transpose of those of
the Koszul-Tate differential $\delta$  
except for the bilinear terms of the ghost momenta 
in eqs. (\ref{eqn:del})~\cite{ht}.
The transformation property of the longitudinal differential $d$ for 
the higher ghost number ghost fields can be obtained by imposing 
$(d\Delta + \Delta d)\eta_n = 0$ and 
$(d\Delta + \Delta d)\eta'_n = 0$ iteratively.
To be more specific we use eqs.(\ref{eqn:ld}) and (\ref{eqn:ad}) as starting 
relations, then $D$ is determined order by 
order of the auxiliary degree by requiring the nilpotency of $D$ on the 
constraint surface.
The results coincide with those obtained from 
eqs.(\ref{eqn:BRST}):
\begin{equation}
 \begin{array}{rcl}
  D \omega_1 & = & d \omega_1 = D_1 \eta_1^F 
                                  - \{ \pi_{\omega_1} , \eta_1^{'F} \}, \\ 
  D \pi_{\omega_1} & = & d \pi_{\omega_1} = [ \pi_{\omega_1} , \eta_1^F ], \\ 
  D \eta_{2n}^B & = & [ \pi_{\omega_1} , \eta_{2n+1}^F ] + 
                   \displaystyle{\sum_{m=1}^{n}}
                   [ \eta_{2m}^B , \eta_{2(n-m)+1}^F ], \\ 
  D \eta_{2n-1}^F & = & \{\pi_{\omega_1} , \eta_{2n}^B \}
               + {\displaystyle{\frac {1}{2} \sum_{m=1}^{n-1}}} 
                       \{ \eta_{2m}^B , \eta_{2(n-m)}^B \}
               + {\displaystyle{\frac {1}{2} \sum_{m=0}^{n-1}}} 
                         \{ \eta_{2m+1}^F , \eta_{2(n-m)-1}^F \}, \\
  D \eta_{2n}^{'B} & = & D_1 \eta_{2n+1}^F 
                            - \{ \pi_{\omega_1} , \eta_{2n+1}^{'F} \} \\
                & &    + {\displaystyle{\sum_{m=1}^{n} \left( 
                          [ \eta_{2m}^{'B} , \eta_{2(n-m)+1}^F ]
                          - \{ \eta_{2m}^B , \eta_{2(n-m)+1}^{'F} \} 
                                                                \right)}}, \\ 
  D \eta_{2n-1}^{'F} & = & D_1 \eta_{2n}^B 
                             - [ \pi_{\omega_1} , \eta_{2n}^{'B} ]
                + {\displaystyle{\sum_{m=1}^{n-1}}}
                   [ \eta_{2m}^{'B} , \eta_{2(n-m)}^B ]
                + {\displaystyle{\sum_{m=0}^{n-1}}}
                   \{ \eta_{2m+1}^{'F} , \eta_{2(n-m)-1}^F \}.
\label{eqn:D} 
 \end{array}
\end{equation}
Then the actions of $D$ for the ghost momenta and of $\stackrel{(k)}{s}$ for 
all fields are determined order by order of the antighost number by requiring 
the nilpotency of the BRST differential $s$.
It is convincing that the result obtained by these procedures leads to 
the BRST differential (\ref{eqn:sham}) which is nilpotent by its 
construction.
For completeness we give the action of $D$ for the ghost momenta and of 
$\stackrel{(k)}{s}$ for all fields in the appendix.

The extended phase space is defined to include the ghosts and ghost momenta 
with a canonical structure
\begin{eqnarray}
 & & [ \pi_{\omega_1} , \omega_1 ]_P = -1, \nonumber\\
 & & [ P_{2m}^B , \eta_{2n}^B ]_P = 
     [ P_{2m-1}^F , \eta_{2n-1}^F ]_P = 
     - \delta_{mn}, \\
 & & [ P_{2m}^{'B} , \eta_{2n}^{'B} ]_P = 
     [ P_{2m-1}^{'F} , \eta_{2n-1}^{'F} ]_P = 
     - \delta_{mn}, \nonumber
\end{eqnarray}
where $[ \ , \ ]_P$ represents the graded Poisson bracket which will be 
replaced by the graded commutation relation 
\Big($[A,B]_P \rightarrow AB-(-)^{|A||B|}BA$, with $|A|,|B|$  Grassmann 
parity of the 
fields $A$ and $B$\Big) multiplied by $-i$ 
upon the quantization, as usual.
Here we have not explicitly shown the coordinate dependence. 
For example the first relation in the above actually means 
$
 [ \pi_{\omega_1}(x_0,x_1) , \omega_1(y_0,y_1) ]_P\mid_{x_0=y_0} 
 = -\delta (x_1-y_1)
$.
Hereafter we omit the coordinate dependence in the Poisson bracket relation. 
These equations can be written in the compact forms
\begin{equation}
  [ \Pi_{2m}^B , \Gamma_{2n}^B ]_P = 
            [ \Pi_{2m-1}^F , \Gamma_{2n+1}^F ]_P = 
             - \delta_{m+n,0}.
\end{equation}
By using this canonical relation, the nilpotent BRST charge $\Omega_{min}$, 
which realizes $sX = [ X , \Omega_{min} ]_P$ 
for $X = \Pi_{2m}^B$, $\Pi_{2m-1}^F$, $\Gamma_{2n}^B$ and $\Gamma_{2n+1}^F$, 
is defined by 
\begin{eqnarray}
  \Omega_{min} & = & \int \mbox{Tr}_{\mbox{\bf 1}}^1 \bigg( \ 
                      \frac{1}{2} \widetilde{\cal A} Q' \widetilde{\cal A} 
                      + \frac{1}{3} {\widetilde{\cal A}}^3   \ \bigg) 
\label{eqn:om} \\
               & = & \int \mbox{Tr}^1 \bigg\{ \ 
                       \hat{A} \left( d_1 \hat{\psi} + 
                       [ A , \hat{\psi} ] \right)
                       - \psi \left( d_1A + A^2 +
                       \hat{\psi}^2 + \hat{A}^2 \right) + 
                       \frac{1}{3}\psi^3\ \bigg\} \nonumber \\
               & = & \int dx^1 \mbox{Tr} \sum_{n=-\infty}^{\infty} \left\{
                      \Pi_{2n}^B \bigg( \partial_1 \Pi_{-2n+1}^F + 
                      \sum_{m=-\infty}^{\infty} \left[ \Gamma_{2m}^B , 
                         \Pi_{-2(m+n)+1}^F \right] \bigg) \right. \nonumber \\
               & & \left. \hspace{2cm}  
                        - \Gamma_{2n+1}^F \sum_{m=-\infty}^{\infty} \left( 
                          \Pi_{2m-1}^F \Pi_{-2(m+n)+1}^F 
                         + \Pi_{2m}^B \Pi_{-2(m+n)}^B 
                                               \right) \right\}, \nonumber 
\end{eqnarray}
where the integration is performed on the one-dimensional space 
and $d_1=dx^1\partial_1$. 
In one dimensional expression in the above, the terms $d_1A, A^2, \psi^3$ 
vanish. 
The upper index 1 on the Tr indicates to pick up only the part with ghost 
number 1 and the subscript ${\bf 1}$ in eq.(\ref{eqn:om}) is necessary 
in order that the integrant contains only fermionic odd forms.
It is very interesting to note that the BRST charge $\Omega_{min}$ 
in the minimal sector
is the fermionic counterpart of the same generalized Chern-Simons action 
in one dimension, as we can compare the above expression with 
$S^f_o$ in (\ref{eqn:comexp}).

In order to fix the gauge, we have to extend the phase space further.
Since we want to make the Landau type gauge-fixing as in the Lagrangian 
formulation we introduce the following set of canonical variables and momenta: 
\begin{eqnarray*}
& & \lambda^B_{2(n-1)}, \ \ \lambda^F_{2n-1}, \ \ 
     b^B_{2(n-1)}, \ \ b^F_{2n-1}, \ \ 
    \bar{C}^B_{2n}, \ \ \bar{C}^F_{2n-1}, \ \ 
    \rho^B_{2n}, \ \ \rho^F_{2n-1},  \\
& & \lambda^{'B}_{2(n-1)}, \ \ \lambda^{'F}_{2n-1}, \ \ 
     b^{'B}_{2(n-1)}, \ \ b^{'F}_{2n-1}, \ \ 
    \bar{C}^{'B}_{2n}, \ \ \bar{C}^{'F}_{2n-1}, \ \ 
    \rho^{'B}_{2n}, \ \ \rho^{'F}_{2n-1},  \\
& & \lambda^{''B}_{2(n-1)}, \ \ \lambda^{''F}_{2n-1}, \ \ 
     b^{''B}_{2(n-1)}, \ \ b^{''F}_{2n-1}, \ \ 
    \bar{C}^{''B}_{2n}, \ \ \bar{C}^{''F}_{2n-1}, \ \ 
    \rho^{''B}_{2n}, \ \ \rho^{''F}_{2n-1},  
            \hspace{1cm} n = 1,2,3,\cdots,
\end{eqnarray*}
where the ghost number of $\lambda_n$ and $\rho_n$ is $n$ while 
that of $\bar{C}_n$ and $b_n$ is $-n$.
The statistics of the even (odd) ghost number fields is bosonic (fermionic).
The canonical structure is defined by 
\begin{equation}
\begin{array}{ccc}
\hfill {[ \rho^B_{2m} , \bar{C}^B_{2n} ]}_P 
   &=& - \delta_{mn}, ~~~~~ 
 {[ b^B_{2m} , \lambda^B_{2n} ]}_P 
   = - \delta_{mn}, \hfill\\
 {[ \rho^F_{2m-1} , \bar{C}^F_{2n-1} ]}_P
   &=& - \delta_{mn}, ~~~~~
 {[ b^F_{2m-1} , \lambda^F_{2n-1} ]}_P 
   = - \delta_{mn}, 
\end{array}
\end{equation}
and the similar canonical structure is defined for the primed fields.
The action of the BRST differential is also extended to these variables as 
\begin{equation}
 \begin{array}{c}
      s \lambda^B_{2n} = \rho^F_{2n+1}, \ \ \ s \rho^F_{2n+1} = 0, \ \ 
      s \bar{C}^F_{2n-1} =  - b^B_{2n}, \ \ \ s b^B_{2n} = 0, \\
      s \lambda^F_{2n-1} = \rho^B_{2n}, \ \ \ s \rho^B_{2n} = 0, \ \ 
      s \bar{C}^B_{2n} =  - b^F_{2n+1}, \ \ \ s b^F_{2n+1} = 0, 
 \end{array} 
\end{equation}
and those for the primed fields are defined in the same way.
The corresponding extended BRST charge is given by 
\begin{eqnarray}
  \Omega &=& \Omega_{min} + \Omega_{nonmin}, \\
  \Omega_{nonmin} &=& \int dx^1 \mbox{Tr} 
           \sum_{n=1}^{\infty} \Big(
             b^B_{2(n-1)} \rho^F_{2n-1} + b^{'B}_{2(n-1)} \rho^{'F}_{2n-1} 
                       + b^{''B}_{2(n-1)} \rho^{''F}_{2n-1} \nonumber\\
  & & \hspace{4cm}       
           - b^F_{2n-1} \rho^B_{2n} - b^{'F}_{2n-1} \rho^{'B}_{2n} 
                                - b^{''F}_{2n-1} \rho^{''B}_{2n} \Big). 
\end{eqnarray}

Now the gauge-fixed action $S$ is obtained by a Legendre transformation from 
the Hamiltonian in the extended phase space:
\begin{eqnarray}
  S & = & \int dx^0 \Bigg\{ \int dx^1 \mbox{Tr} \Big(
        \dot{\omega}_1 \pi_{\omega_1}  + \sum_{n=1}^{\infty} \big( \ 
               \dot{\eta}^B_{2n} P^B_{2n} 
             + \dot{\eta}^F_{2n-1} P^F_{2n-1}
             + \dot{\eta'}^{B}_{2n} P^{'B}_{2n} 
             + \dot{\eta'}^{F}_{2n-1} P^{'F}_{2n-1} \nonumber \\
       & &  \hspace{1.5cm}
             + \dot{\lambda}^B_{2(n-1)} b^B_{2(n-1)} 
             + \dot{\lambda}^F_{2n-1} b^F_{2n-1}
             + \dot{\lambda}^{'B}_{2(n-1)} b^{'B}_{2(n-1)}  
             + \dot{\lambda}^{'F}_{2n-1} b^{'F}_{2n-1}        \nonumber \\
       & &  \hspace{1.5cm} 
             + \dot{\lambda}^{''B}_{2(n-1)} b^{''B}_{2(n-1)}  
             + \dot{\lambda}^{''F}_{2n-1} b^{''F}_{2n-1}
             + \dot{\bar{C}}^B_{2n} \rho^B_{2n} 
             + \dot{\bar{C}}^F_{2n-1} \rho^F_{2n-1}  \nonumber \\
       & &  \hspace{1.5cm} 
             + \dot{\bar{C}'}^{B}_{2n} \rho^{'B}_{2n} 
             + \dot{\bar{C}'}^{F}_{2n-1} \rho^{'F}_{2n-1} 
             + \dot{\bar{C}''}^{B}_{2n} \rho^{''B}_{2n} 
             + \dot{\bar{C}''}^{F}_{2n-1} \rho^{''F}_{2n-1} 
                 \ \big) \Big)  - H_K \Bigg\}, 
\label{eqn:aham} \\
      & & H_K = [ \ K \ , \ \Omega \ ]_P, 
\label{eqn:kham}           
\end{eqnarray}
where $K$ is called a gauge-fixing fermion.
The gauge-fixed Hamiltonian $H_K$ consists of gauge-fixing and ghost parts 
only since the total Hamiltonian of the system have vanished.
There is no systematic way to find $K$ so as to yield a covariant expression.
Actually we want to show that the action obtained in the Hamiltonian 
formulation coincides with that in the Lagrangian formulation.
We take the following gauge-fixing fermion $K$ based on the experience 
of the quantization of $gl(1,{\bf R})$ model:
\begin{eqnarray}
  K & = & \int dx^1 \mbox{Tr} \sum_{n=1}^{\infty} \Big\{ 
           \epsilon_1 \bar{C}^F_{2n-1} \partial^1 \eta'^{B}_{2(n-1)}  
         + \epsilon_2 \bar{C}^B_{2n} \partial^1 \eta'^{F}_{2n-1} 
         + \epsilon_3 \bar{C}^{'F}_{2n-1} \partial_1 \lambda^{''B}_{2(n-1)} 
\nonumber\\
   & &   + \epsilon_4 \bar{C}^{'B}_{2n} \partial_1 \lambda^{''F}_{2n-1} 
         + \epsilon_5 \bar{C}^{''F}_{2n-1} \partial^1 \lambda^{'B}_{2(n-1)} 
         + \epsilon_6 \bar{C}^{''B}_{2n} \partial^1 \lambda^{'F}_{2n-1} 
\nonumber\\
   & &   + \epsilon_7 P^F_{2n-1} \lambda^B_{2(n-1)} 
         + \epsilon_8 P^B_{2n} \lambda^F_{2n-1} 
         + \epsilon_9 P^{'F}_{2n-1} \lambda^{'B}_{2(n-1)} 
         + \epsilon_{10} P^{'B}_{2n} \lambda^{'F}_{2n-1} \Big\},
\end{eqnarray}
where we denote $\omega_1$ by $\eta'_0$ and the sign factors 
$\epsilon_i = \pm 1 ( i = 1, 2, \cdots, 10 )$ will be determined soon.

We first impose one gauge-fixing condition, 
which is originated from the terms 
$\bar{C}^{B,F}_n \partial^1 \eta'^{F,B}_{n-1}$ in $K$,
at each level of the ghost number in the minimal sector.
It is consistent with the fact that the number of the linearly independent 
constraints or reducibility conditions at each level 
should be one, as mentioned above.
Next we substitute this gauge-fixing fermion into eqs.(\ref{eqn:aham}) and 
(\ref{eqn:kham}) and integrate out the momentum variables 
$P^{B,F}_n$, $P^{'B,F}_n$, $\rho^{B,F}_n$ and $\rho'^{B,F}_n$.
If we set 
$$
\epsilon_2 = \epsilon_3 = \epsilon_7 = \epsilon_8 = 1, \qquad 
\epsilon_1 = \epsilon_5 = -1, \qquad
\epsilon_6 = - \epsilon_4, 
$$
and rename the variables as
\begin{eqnarray*}
  \pi_{\omega_1} & \longrightarrow &  - \phi, \\ 
  ( \eta^B_{2n}, \ \eta^F_{2n+1}, \ \eta'^B_{2n}, \ \eta'^F_{2n+1} )  
                & \longrightarrow & 
        ( -C^B_{2n}, \ C^F_{2n+1}, \ C^B_{2n \mu=1}, \ -C^F_{2n-1 \mu=1} ), \\
  ( \lambda^B_{2n}, \ \lambda^F_{2n+1}) 
                & \longrightarrow &  
        ( - C^B_{2n \mu=0}, \ - C^F_{2n-1 \mu=0} ), \\
  ( \lambda'^B_{2n}, \ \lambda'^F_{2n+1} ) 
                & \longrightarrow &
        ( \epsilon_9 \widetilde{C}^B_{2n}, \ 
                      \epsilon_{10} \widetilde{C}^F_{2n+1} ), \\
  ( \lambda''^B_{2n}, \ \lambda''^F_{2n+1} ) 
                & \longrightarrow &
        ( \epsilon_9 \eta^B_{2n}, \ 
                      \epsilon_4 \epsilon_{10} \eta^F_{2n+1} ), \\ 
  ( b^B_{2n}, \ b^F_{2n+1} ) 
                & \longrightarrow &
        ( -b^B_{2n}, \ -b^F_{2n+1} ), \\
  ( b'^B_{2n}, \ b'^F_{2n+1} ) 
                & \longrightarrow & 
        ( \epsilon_9 b_{2n}^{B \mu=1}, \ 
                      \epsilon_{10} b_{2n+1}^{F \mu=1} ), \\
  ( b''^B_{2n}, \ b''^F_{2n+1} ) 
                & \longrightarrow & 
        ( \epsilon_9 b_{2n}^{B \mu=0}, \ 
                     \epsilon_4 \epsilon_{10} b_{2n+1}^{F \mu=0}), \\
  ( \bar{C}'^B_{2n}, \ \bar{C}'^F_{2n+1} ) 
                & \longrightarrow & 
        ( - \epsilon_{10} \bar{C}_{2n}^{B \mu=1}, \ 
                                   - \epsilon_9 \bar{C}_{2n+1}^{F \mu=1} ), \\ 
  ( \bar{C}''^B_{2n}, \ \bar{C}''^F_{2n+1} ) 
                & \longrightarrow & 
        ( - \epsilon_4 \epsilon_{10} \bar{C}_{2n}^{B \mu=0}, \ 
                              - \epsilon_9 \bar{C}_{2n+1}^{F \mu=0} ), \\
  ( \rho''^B_{2n}, \ \rho''^F_{2n+1} ) 
                & \longrightarrow &
        ( \epsilon_4 \epsilon_{10} \pi^B_{2n}, \ 
                         \epsilon_9 \pi^F_{2n+1} ), \\
  C^B_{n=0 \ \mu=0} = \omega_0, & & \widetilde{C}^B_0 = B,
\end{eqnarray*}
this action completely coincides with the gauge-fixed action (\ref{eqn:gfa}) 
in the Lagrangian formulation in which the propagators of all fields are 
well-defined\footnote{
%%%%%%footnote%%%%%%%
Three sign factors $\epsilon_4, \epsilon_9, \epsilon_{10}$ have remained 
arbitrary.
%%%%%%%%%%%%%%%%%%%%%
}.
This result gives an evidence to the statement that the number of linearly 
independent constraints between (\ref{eqn:g1}) and (\ref{eqn:h1}) 
should be one.

%%%%%%%%%%%%%%%%%%%%
\subsection{Inclusion of fermions in the Hamiltonian formulation}

In the case where the classical action contains fermionic fields in addition 
to bosonic fields we can follow the similar procedure as the one carried out 
in the previous subsection.
We explain how the analyses given in the purely bosonic model 
must be modified. 

First we obtain the constraints
\begin{eqnarray}
  \mbox{second class} \quad  
    & & \pi_{\phi} = 0, \ \ \ \pi_{\omega_1} + \phi = 0, 
\label{eqn:secb} \\
    & & \pi_{\chi} = 0, \ \ \ \pi_{\chi_1} + \chi = 0, 
\label{eqn:secf} \\
  \mbox{first class} \quad 
    & & \pi_{\omega_0} = 0, \ \ \ \pi_B = 0, 
\label{eqn:fir1} \\
    & & \pi_{\chi_0} = 0, \ \ \ \pi_{\tilde{\chi}} = 0, 
\label{eqn:fir2} \\
    & & D_1 \pi_{\omega_1} + [ \pi_{\phi} , \pi_{\omega_1} ] 
               + \{ \pi_{\chi_1} , \chi_1 - \pi_{\chi} \} = 0, \\
    & & \pi_{\omega_1}^2 + \pi_{\chi_1}^2 = 0, \\
    & & D_1 \pi_{\chi_1} + [ \pi_{\phi} , \pi_{\chi_1} ] 
               - \{ \pi_{\omega_1} , \chi_1 - \pi_{\chi} \} = 0, \\
    & & [ \pi_{\omega_1} , \pi_{\chi_1} ] = 0,
\end{eqnarray}
and the total Hamiltonian
\begin{eqnarray*}
  H_T = \int dx^1 \mbox{Tr} 
             &&  \Big\{ 
                 \omega_0 \Big( -D_1 \pi_{\omega_1} - [ \pi , \pi_{\phi} ] 
                             -\{ \chi , \pi_{\chi} \} 
                             -\{\chi_1 , \pi_{\chi_1} \} \Big) \\
             && 
                + B \Big( -\phi^2 - \chi^2 - \{ \pi_{\omega_1} , \phi \}
                         - \{ \chi , \pi_{\chi_1} \} \Big) \\
             && 
               + \chi_0 \Big( -D_1 \pi_{\chi_1} + [ \pi_{\phi} , \chi ]
                             +  \{ \pi_{\omega_1} , \chi_1 \}
                             + \{ \phi , \pi_{\chi} \} \Big) \\
             && 
                  + \tilde{\chi} \Big( [ \phi , \chi ] + [ \pi_{\phi} , \chi ]
                                    + \{ \pi_{\omega_1} , \chi_1 \}
                                    + \{ \phi , \pi_{\chi} \} \Big) \\
             && 
                 + \lambda_{\omega_0} \pi_{\omega_0} 
                 + \lambda_B \pi_B + \lambda_{\chi_0} \pi_{\chi_0} 
                 + \lambda_{\tilde{\chi}} \pi_{\tilde{\chi}} \Big\}.
\end{eqnarray*}
Then by using Dirac's brackets for the second class constraints 
(\ref{eqn:secb}) and (\ref{eqn:secf}) and adopting gauge conditions 
$\omega_0 = B = \chi_0 = \tilde{\chi} = 0$ for the first class constraints 
(\ref{eqn:fir1}) and (\ref{eqn:fir2}) we have four phase space variables 
$\omega_1$, $\pi_{\omega_1}$, $\chi_1$ and $\pi_{\chi_1}$ 
with the first class constraints
\begin{eqnarray}
  G^B_1 &\equiv& -D_1 \pi_{\omega_1} - \{ \pi_{\chi_1} , \chi_1 \} = 0,\\
  H^B_1 &\equiv& - \pi_{\omega_1}^2 - \pi_{\chi_1}^2 = 0,\\
  G^F_1 &\equiv& D_1 \pi_{\chi_1} - \{ \pi_{\omega_1} , \chi_1 \} = 0,
\label{eqn:firf1}\\
  H^F_1 &\equiv& - [ \pi_{\omega_1} , \pi_{\chi_1} ] = 0, 
\label{eqn:firf2}
\end{eqnarray}
and now the total Hamiltonian vanishes.
It is understood that these constraints are infinitely reducible 
due to the following relations which hold on the constraint surface:
\begin{equation}
\begin{array}{rcl}
  G^B_n &\equiv& (-)^{n+1} D_1 H^B_{n-1} 
                  + [ \pi_{\omega_1} , G^B_{n-1} ]_{(-)^n} \\
      & &        + (-)^{n+1} \{ \pi_{\chi_1} , G^F_{n-1} \} 
                 + [ \chi_1 , H^F_{n-1} ]_{(-)^{n+1}} = 0, \\
  H^B_n &\equiv& [ \pi_{\omega_1} , H^B_{n-1} ]_{(-)^{n+1}}
                + (-)^{n+1} \{ \pi_{\chi_1} , H^F_{n-1} \} = 0, \\
  G^F_n &\equiv& (-)^n D_1 H^F_{n-1} 
                   + (-)^n [ \pi_{\chi_1} , G^B_{n-1} ] \\
      & &       + [ \chi_1 , H^B_{n-1} ]_{(-)^{n+1}}
                + [ \pi_{\omega_1} , G^F_{n-1} ]_{(-)^{n+1}} = 0, \\
  H^F_n &\equiv& (-)^n [ \pi_{\chi_1} , H^B_{n-1} ] 
                + [ \pi_{\omega_1} , H^F_{n-1} ]_{(-)^n} = 0,
                   \hspace{2cm} n = 2,3,4,\cdots.
\end{array}
\end{equation}
Next we further introduce infinite ghosts and ghost momenta 
in addition to those of (\ref{eqn:gh}) in the purely bosonic model:
$$
 \begin{array}{llllllll}
  \eta_{2n}^F, && \eta'^{F}_{2n}, 
               && P_{2n}^F, && P'^{F}_{2n}, & \\
 \eta_{2n-1}^B, && \eta'^{B}_{2n-1}, 
                && P_{2n-1}^B, && P'^{B}_{2n-1}, 
                &\hspace{2cm} n = 1,2,3,\cdots.
 \end{array}
$$
Since these fields correspond to the fermionic constraints (\ref{eqn:firf1}) 
and (\ref{eqn:firf2}) the fields with odd (even) ghost numbers are bosonic 
(fermionic).
Thus we now have $\eta^{B,F}_n$, $\eta'^{B,F}_n$, $P^{B,F}_n$ 
and $P'^{B,F}_n$ for all ghost numbers.
The system we now consider possesses the same algebraic structure as that of 
the purely bosonic model because in the generalized Chern-Simons theories 
fermionic and bosonic fields are treated in a unified manner.
Therefore what we have to do is just to replace indices of the ghosts and 
the ghost momenta which were either even or odd integers 
in the purely bosonic model to all integers.
After these modifications we can show that the same BRST transformations 
and the gauge-fixed action as those in the Lagrangian formulation 
are obtained in the Hamiltonian formulation.

%%%%%%%%%%%%%%%%%%%%
\section{Perturbative aspects of the model} 

\setcounter{equation}{0}
\setcounter{footnote}{0}

%%%%%
\subsection{Partition function}

In this section we present a perturbative analysis of the quantized 
gauge-fixed action (\ref{eqn:gfa}). 
First we investigate the partition function of the model.
It is expected that the partition function is simply equal to 1 
due to the topological nature of the model,
which can be also understood in the Hamiltonian formulation where
it is shown that there is no local physical dynamical variables.
We show in the following how it is realized in a 
certain regularization scheme.

It is easy to see that the partition function is one loop exact,
which will be proved in the next subsection by treating higher order
corrections generally. 
In this subsection we evaluate the one loop contribution
to the partition function.
We have only to extract the
quadratic terms from the gauge-fixed action and evaluate the determinant
factors coming from the Gaussian path-integrations. In the case of the model
containing both bosonic and fermionic fields in the starting classical action
treated in section~2 and 3, the bosonic and fermionic fields possess the
same kinetic terms (\ref{eqn:totkt}) up to the sign factors
due to the Grassmannian property 
of fermionic fields. Therefore the partition function is trivially 1 since
the contribution from the bosonic and fermionic fields cancels in each 
ghost number sector. Although this can be attributed to a (scalar) 
supersymmetry of the kinetic terms, it is not a symmetry of the full
gauge-fixed action. 
Thus the topological nature behind the triviality 
of the partition function is hidden by this symmetry of the kinetic term. 
On the other hand, in the case of the purely bosonic model
treated in the previous paper, the triviality follows from the cancellation
among infinite ghost number sectors, which we show in the following.

The quadratic terms of the gauge-fixed action of the purely bosonic model
are
\begin{eqnarray*}
  S_{kin} & = & \int d^2 x {\mbox{Tr}} \Big\{ \ 
               - \phi \epsilon^{\mu \nu} \partial_{\mu} \omega_{\nu} 
              +  \partial^{\mu} \omega_{\mu} b_0 + 
               \epsilon_{\mu \nu}^{-1} \partial^{\nu} B \ b_0^{\mu}   
              + \partial_{\mu} \eta_{0} \  b_{0}^{\mu} \\
          & &  \ \ \ \ \ \ \ \ \ \ \ \  \ \  
                + \sum_{n=1}^{\infty} \Big( 
                  - \partial^{\mu} \bar{C}_n \partial_{\mu} C_n
                  - \frac{1}{2} ( \partial^{\mu} \bar{C}_n^{\nu} 
                                 - \partial^{\nu} \bar{C}_n^{\mu} ) 
                                ( \partial_{\mu} C_{n \nu} 
                                 - \partial_{\nu} C_{n \mu} ) \Big)  \\
        & & \ \ \ \ \ \ \ \ \ \ \ \ \ \ 
               +  \sum_{n=1}^{\infty} \left( 
                \partial^{\mu} C_{n \mu} b_{n} + 
                \epsilon_{\mu \nu}^{-1} \partial^{\nu} 
                        \widetilde{C}_{n} b_n^{\mu}
                      + \partial_{\mu} \eta_{n-1}  b_{n-1}^{\mu} - 
               \partial_{\mu} \bar{C}_n^{\mu} \pi_n \right) \Big\},
\end{eqnarray*}
where we adopt the notation in the previous paper and the fields of
odd (even) ghost number are fermionic (bosonic).
It is straightforward to find the contributions to the partition
function $Z$ from each ghost number sector.
Path-integral over fields of ghost number $\pm n \ne 0$ gives
$(\det \triangle^{(0)})^{4\epsilon_n}$ where $\epsilon_n$ is $+1$ ($-1$)
for odd $n$ (even $n$) while fields of ghost number 0 contribute
$(\det \triangle^{(0)})^{-2}$. Here $\triangle^{(0)}$ is the Laplacian
for zero forms and we have used the relation of the Laplacian for 
one forms $\det \triangle^{(1)}=(\det \triangle^{(0)})^2$.
Thus the log of the partition function becomes
\begin{eqnarray}
  \ln Z = - 4 \ln \det \triangle^{(0)} \times 
            \Bigl( {1\over2}+\sum_{n=1}^\infty
                    (-)^n \Bigr).  
\label{eqn:lnz}
\end{eqnarray}
The contribution from the ghost number zero sector is twice overcancelled
by that from the ghost number $\pm 1$ sector and the sum is again
overcancelled by that from the ghost number $\pm 2$ sector, and  so on.
This reflects the structure of the reducibility of the model that
the number of gauge and reducibility parameters are twice the
real gauge degrees of freedom which can actually gauge away
all local dynamical variables. The gauge-fixed action is defined
as an infinite series which possesses the BRST invariance, which implies
that the contribution from the ghost number zero sector should be
canceled by the sum of contributions from nonzero ghost number sectors.
Therefore we should require that the sum in eq.(\ref{eqn:lnz}) be zero
as a regularization for the summation over infinite ghost number sectors. 
This can be accomplished by the zeta function regularization, which
leads to $\displaystyle \sum_{n=1}^\infty (-)^n= -\frac{1}{2}$. 
We adopt the zeta function regularization
henceforth as a regularization for the summation of infinite series of
ghost contributions.

It should be noted that though we have discussed the triviality of the
partition function in the space with flat metric, the same arguments
hold in curved spaces except for the contribution from the zero mode
due to the global structure of the space.

%%%%%%%%%%
\subsection{Higher order corrections}

In this subsection we investigate loop effects of the gauge-fixed action
(\ref{eqn:gfa}). It is convenient to path-integrate out the auxiliary fields 
$b_n$, $\eta_n$ and $\pi_n$, which imposes the Landau gauge conditions
\begin{equation}
 \begin{array}{ccc}
     \partial^\mu C_{n\mu}^F=0, & \quad \partial_\mu b_n^{F\mu}=0, 
                                & \quad \partial_\mu \bar{C}_n^{F\mu}=0, \\
     \partial^\mu C_{n\mu}^B=0, & \quad \partial_\mu b_n^{B\mu}=0,
                                & \quad \partial_\mu \bar{C}_n^{B\mu}=0. 
\label{eqn:lg}  
 \end{array} 
\end{equation}
We introduce the following compact notation
\begin{equation}
\begin{array}{rclcrcl}
  C^F & = & \displaystyle{\sum_{n=-\infty}^\infty} C_n^F,& \quad &
        C^B & = & \displaystyle{\sum_{n=-\infty}^\infty} C_n^B,  \cropen
  C_{\mu}^F & = & \displaystyle{\sum_{n=-\infty}^\infty} C_{\mu n}^F, & \quad &
        C_\mu^B & = & \displaystyle{\sum_{n=-\infty}^\infty}
                                                     C_{\mu n}^B,  \cropen
  \Ctilde^F & = & \displaystyle{\sum_{n=-\infty}^\infty} \Ctilde_n^F,& \quad &
        \Ctilde^B & = & \displaystyle{\sum_{n=-\infty}^\infty} \Ctilde_n^B,
                                                                   \cropen
  b^{F\mu} & = & \displaystyle{\sum_{n=0}^\infty} \hspace*{2mm}
                                                      b_n^{F\mu},& \quad &
  b^{B\mu} & = & \displaystyle{\sum_{n=0}^\infty} \hspace*{2mm} b_n^{B\mu}. 
\end{array}
\end{equation}
Here the fields with negative ghost number indices are 
\begin{equation}
 \begin{array}{rclcrclr}
 C_{-n}^F &=& \epsilon^{-1}_{\mu\nu}\partial^{\mu}\Cbar^{F\nu}_n, &\quad&
 C_{-n}^B &=& -\epsilon^{-1}_{\mu\nu}\partial^{\mu}\Cbar^{B\nu}_n,& \\
 C_{\mu -n}^F &=& -\epsilon^{-1}_{\mu\nu}\partial^\nu \Cbar_{n}^F, &\quad&
 C_{\mu -n}^B &=& -\epsilon^{-1}_{\mu\nu}\partial^\nu \Cbar_{n}^B, & \\
 \Ctilde_{-n}^F &=& 0, &\quad& \Ctilde_{-n}^B &=& 0, 
 & \hspace{2cm} n=1, 2, 3, \cdots,
 \end{array}
\end{equation}
where the identifications (\ref{eqn:idantif}) and (\ref{eqn:ioat}) are used.
Then the gauge-fixed action (\ref{eqn:gfa}) is rewritten in the form
\begin{eqnarray}
        S_{tot}&=&-\int d^2x\, \Gh^0\Bigl\{\eta_{ab} 
        \epsilon^{\mu\nu}\Bigl(C_\mu^{Ba}\psnu C^{Bb}
                                  +C_\mu^{Fa}\psnu C^{Fb}\Bigr)
        +\eta_{ab}\epsilon_{\mu\nu}^{-1}\Bigl(
         \pnu\Ctilde^{Ba}b^{Bb\mu}
        + \pnu\Ctilde^{Fa}b^{Fb\mu}\Bigr) \nonumber\\
    & & \quad+F_{abc}\Bigl(
      {1\over2}\epsilon^{\mu\nu}C_\mu^{Ba}C_\nu^{Bb}C^{Bc}
      +\epsilon^{\mu\nu}C_\mu^{Fa}C_\nu^{Bb}C^{Fc}
      +C^{Ba}C^{Fb}\Ctilde^{Fc}+{1\over2}C^{Fa}C^{Fb}\Ctilde^{Bc}\Bigr) 
      \nonumber\\
    & &\quad +D_{abc}\Bigl(
      -{1\over2}\epsilon^{\mu\nu}C_\mu^{Fa}C_\nu^{Fb}C^{Bc}
      +{1\over2}C^{Ba}C^{Bb}\Ctilde^{Bc}\Bigr)\Bigr\},
\end{eqnarray}
where 
$F_{abc}={\mbox {Tr}}(T_a[T_b,T_c]) = \eta_{ad} f^d_{bc}$ 
is a totally antisymmetric structure constant while 
$D_{abc}={\mbox {Tr}}(T_a\{T_b,T_c\}) = \eta_{ad} h^d_{bc}$ 
is a totally symmetric structure constant.
$\Gh^0$ represents to pick up the zero ghost
number terms as in eq.(\ref{eqn:MA}). 
Since the field $b^{B\mu}$ and $b^{F\mu}$ do not interact with other fields, 
the perturbation theory is much simplified.

From this expression and the Landau gauge conditions (\ref{eqn:lg}), 
we can obtain propagators\footnote{
%%%%%footnote%%%%%
        We take the flat Minkowski metric $g_{\mu\nu}=\hbox{diag}(-1,+1)$.
        The propagators are thus obtained by the Minkowskian path-integral.
        }
%%%%%%%%%%%%%%%%%%
%
\begin{eqnarray}
\langle C^{Ga}_{-n}(x) C^{Gb}_{m\mu}(0)  \rangle & = & 
        \int {d^2p\over (2\pi)^2} e^{ipx} 
             {-\epsilon_{\mu\nu}^{-1}\, p^\nu \over p^2}
                \delta_{nm}(\eta^{-1})^{ab}
        \equiv D_\mu(x)\delta_{nm}(\eta^{-1})^{ab},  \nonumber\\
\langle b^{Ga\nu}_{n}(x) \Ctilde^{Gb}_{m}(0) \rangle & = & 
        \int{d^2p\over (2\pi)^2} e^{ipx} 
           {-\epsilon^{\mu\nu}\, p_\nu \over p^2}
                \delta_{nm}(\eta^{-1})^{ab}       
        \equiv \Dtilde^\mu(x)\delta_{nm}(\eta^{-1})^{ab}, \\
        && \hspace{8cm} G=B, F, \nonumber
\end{eqnarray}
which imply the propagators for the component fields:
\begin{eqnarray*}
\langle \phi^a(x) \omega_\mu^b(0)\rangle &=& 
   \langle \chi^a(x) \chi_\mu^b(0)\rangle = D_\mu(x) (\eta^{-1})^{ab},   \\
\langle C^{Ga}_{n\mu}(x) \Cbar^{Gb\nu}_m(0)\rangle &=& 
        - i \int {d^2p\over (2\pi)^2}  e^{ipx}
                \Bigl(\delta^\nu_\mu-{p_\mu p^\nu \over p^2}\Bigr) 
                {1\over p^2} \delta_{nm}(\eta^{-1})^{ab}, \\
\langle C^{Ga}_{n}(x) \Cbar^{Gb}_m(0)\rangle &=& 
        - i \int {d^2p\over (2\pi)^2}  e^{ipx}
                {1\over p^2} \delta_{nm}(\eta^{-1})^{ab}, \\
\langle b^{Ga\mu}_n(x) \Ctilde^{Gb}_m(0)\rangle &=& 
                \Dtilde^\mu(x) \delta_{nm}(\eta^{-1})^{ab}.     
\end{eqnarray*}

We now investigate the effective action obtained as a sum of 1PI diagrams.
For a 1PI graph contributing to the effective action we denote the numbers 
of external legs as $E_C$, $E_{C_\mu}$ and $E_{\Ctilde}$ for
$C^G$, $C_\mu^G$ and $\Ctilde^G$, respectively,\footnote{
%%%%%footnote%%%%%
        We do not discriminate fermionic and bosonic fields here.
        Thus $E_C$ is the sum of the number of external legs for
        $C^F$ and $C^B$, for example.}
%%%%%%%%%%%%%%%%%%
the number of propagators as $P$ and the number of loops as $L$. 
As for vertices,
we classify them into two categories, $C-C_\mu-C_\nu$ type and 
$C-C-\Ctilde$ type, and denote the numbers of vertices
as $V$ and $\Vtilde$, respectively.
Then the following relations hold:
\begin{equation}
\begin{array}{rcl}
E_\Ctilde&=&\Vtilde,  \\
V+2\Vtilde &=& P+E_C, \\
2V &=& P+E_{C_\mu},   \\
L&=&P-(V+\Vtilde)+1.    
\end{array}
\end{equation}
From these relations we obtain
\begin{equation}
L=E_\Ctilde-E_C+1,
\end{equation}
which shows that multiloop graphs must be accompanied by external
legs of $\Ctilde$. In particular the partition function is one loop exact 
as discussed in the previous subsection.
We further obtain the superficial degree of divergence
\begin{equation}
D=2L-P=2-2E_\Ctilde-E_{C_\mu}.
\end{equation}
This implies that the possible ultraviolet divergences exist only for
$L=1$, $E_{C_\mu}=1,2$, $E_C=E_\Ctilde=0$; 
$L=1$, $E_\Ctilde=E_C=1$, $E_{C_\mu}=0$
and $L=2$, $E_\Ctilde=1$, $E_C=E_{C_\mu}=0$ besides the partition function,
which we have shown to be one. In the following we will see that all these
contributions actually vanish in the regularization scheme used
for the partition function and thus the theory is free from 
the ultraviolet divergence.

First we examine one loop diagrams with two external legs. For diagrams
with two bosonic external fields, $C_{m\mu}^{Ba}(x_1)$ and 
$C_{-m\nu}^{Ba'}(x_2)$ or $C_{m}^{Ba}(x_1)$ and $\Ctilde_{-m}^{Ba'}(x_2)$, 
there are two types of graphs: 
those with a loop of bosonic fields and of
fermionic fields. Each graph gives the same contribution 
except for the sign, \ie, 
a fermion loop gives an extra sign factor. Thus bosonic loops and fermionic
loops cancel with each other. We have two comments in order.
First each contribution itself is divergent. For example
a bosonic loop with the first type of external legs yields
\begin{eqnarray}
\int d^2x_1 d^2x_2& & 
        C_{m\mu}^{Ba}(x_1) C_{-m\nu}^{Ba'}(x_2)
        F_{abc}F_{a'b'c'} (\eta^{-1})^{bb'} (\eta^{-1})^{cc'} \nonumber\\
        & & (-i)^2
        \epsilon^{\mu\mu'}\epsilon^{\nu\nu'}
        D_{\mu'}(x_1-x_2) D_{\nu'}(x_2-x_1),
\label{eqn:abc}
\end{eqnarray}
which is logarithmically divergent to be consistent with 
the superficial degree of divergence. 
The same regularization should be applied both for bosonic and
fermionic loops, which makes the whole contribution vanish.
The second comment is for the case of purely bosonic models.
In this case bosonic loops and fermionic loops appear alternatively
according to the ghost number of the fields of the loop.
Thus the cancellation does not work at each fixed ghost number sector
and the whole contribution is $\displaystyle\sum_{n=-\infty}^\infty (-)^n$ 
multiplied by
a bosonic loop contribution of the form (\ref{eqn:abc}). This is reminiscent
of the case for the partition function. We should adopt the zeta function
regularization for the summation to have 
$\displaystyle\sum_{n=-\infty}^\infty(-)^n=0$.
This is closely related with the BRST invariance of the theory.
Indeed if it were not set to zero, the BRST Ward-Takahashi identity
$\langle \delta_B C_{-1}\rangle=0$ would be violated.

For diagrams with two fermionic external fields, $C_{m\mu}^{Fa}(x_1)$ and 
$C_{-m\nu}^{Fa'}(x_2)$ or $C_{m}^{Fa}(x_1)$ and $\Ctilde_{-m}^{Fa'}(x_2)$,
there are two graphs for each ghost number of the loop fields: 
a field of ghost number $n+m$ can be bosonic or fermionic.
These contribution, however, does not cancel with each other. 
For example a diagram with the first type of external fields 
gives the contribution
\begin{eqnarray}
\int d^2x_1& d^2x_2 & 
        C_{m\mu}^{Fa}(x_1) C_{-m\nu}^{Fa'}(x_2)
        F_{abc}D_{a'b'c'} (\eta^{-1})^{bb'} (\eta^{-1})^{cc'} \nonumber\\
        & &(-i)^2
        \epsilon^{\mu\mu'}\epsilon^{\nu\nu'}
        D_{\mu'}(x_1-x_2) D_{\nu'}(x_2-x_1).
\end{eqnarray}
It vanishes, however, by itself due to the totally symmetric and 
antisymmetric property 
of $D_{abc}$ and $F_{abc}$. Thus we have shown one loop diagrams with
two external legs vanish completely.

Next we investigate two loop tadpole diagrams with an external $\Ctilde^G$.
Due to the ghost number conservation and Grassmannian property,
the only possible contribution appears for $\Ctilde_0^B=B$.
It vanishes, however, since the two loop tadpole diagrams contain 
one loop subdiagrams with two external legs which vanish as we have shown
above. Finally the one loop tadpole with an external $C_\mu^G$ vanishes
due to the Lorentz invariance of the loop integration.
Thus we have shown that all possibly divergent diagrams vanish.

Although the generalized Chern-Simons theory is free from the
ultraviolet divergence, there appear infinite quantities to be
taken care of due to the existence of the infinite fields.
As an illuminating example we consider a one loop diagram
with three external legs. For diagrams with external legs
$C_\mu^B C_\nu^B C_\lambda^B$ or $C_\mu^B C^B \Ctilde^B$,
contributions from bosonic loops and fermionic loops cancel with
each other as in the previous case.  For diagrams with external legs
$C_\mu^B C_\nu^F C_\lambda^F$ or $C_\mu^B C^F \Ctilde^F$, however,
cancellations among diagrams or vanishing due to the group theoretic
property do not occur. For example we have
\begin{eqnarray}
\int d^2x_1 d^2x_2& d^2x_3& \   
        C_{m+n\mu}^{Ba_1}(x_1) C_{-m\nu}^{Fa_2}(x_2) C_{-n\lambda}^{Fa_3}(x_3)
           \nonumber \\
         & &(\eta^{-1})^{bb'} (\eta^{-1})^{cc'} (\eta^{-1})^{dd'}
       (F_{b'a_1c}F_{c'a_2d}D_{d'a_3b}-F_{b'a_1c}D_{c'a_2d}F_{d'a_3b})
             \nonumber\\
         & & \epsilon^{\mu\mu'}\epsilon^{\nu\nu'}\epsilon^{\lambda\lambda'}
        D_{\mu'}(x_1-x_2) D_{\nu'}(x_2-x_3) D_{\lambda'}(x_3-x_1)
        \sum_{k=-\infty}^\infty 1,
\end{eqnarray}
as the first type of contribution. 
The summation $\displaystyle\sum_{k=-\infty}^\infty 1$
is ambiguous even using the zeta function regularization since the
summation extends both plus and minus infinity.
This appears to imply that the gauge-fixed action (\ref{eqn:gfa}) is not
fully consistent at the quantum level since we do not have counter terms
to control those divergences. 
However we may be able to regard them as gauge-fixing artifacts.
The point is that the divergent 1PI diagrams contribute only to 
unphysical correlation functions. 
In the present analysis with the flat metric, the only physical
variable is the zero mode of $\phi$ field and other BRST singlet
operators are trivial ones. 
Thus physical correlation 
functions which are independent of gauge-fixing conditions 
are zero and thus free from divergence. Therefore there may exist 
better gauge-fixing 
conditions which yield finite correlation functions also 
for the unphysical sector. Keeping this in mind, we can take
$\displaystyle\sum_{k=-\infty}^\infty 1$ as an integer value $N$ 
by the zeta function regularization, which leaves the value of $N$
arbitrary. Then  physical correlation functions are independent of
$N$ while unphysical ones become finite. It should be noted
that BRST Ward-Takahashi identities are satisfied after this
{\it regularization}. These are the perturbative aspects of
the gauge-fixed action (\ref{eqn:gfa}). It is an open question whether
it gives interesting information even with this sick property
in the unphysical sector.

%%%%%%%%%%%%%%%%%%%%
\section{Quantization of even-dimensional models}

\setcounter{equation}{0}
\setcounter{footnote}{0}

%%%%%%%%%%
\subsection{Four-dimensional model}

The arguments of infinite reducibility given in subsection 3.1 for 
the two-dimensional model are applicable to any even-dimensional models of 
generalized Chern-Simons actions with the minor change of introducing 
higher form gauge parameters. 
In other words the infinite reducibility is a common feature of the 
generalized Chern-Simons actions in any dimensions even including odd 
dimensions.
Furthermore the procedure of constructing the BRST transformation and 
gauge-fixed action is dimension independent and thus applicable 
to any even-dimensional models.
In this subsection we present the quantization of the purely bosonic 
four-dimensional model by the Lagrangian formulation, explicitly, 
and show how the quantization of the four-dimensional model goes 
parallel to the case of the two-dimensional model.

The four-dimensional action without fermionic gauge fields 
is given in eq.(\ref{eqn:4d-act}).
The corresponding gauge transformations without fermionic gauge 
parameters ($\alpha = \hat{\alpha}=0$) are given by 
\begin{eqnarray}
\delta\phi &=& [\phi,v], \nonumber\\
\delta\omega &=& dv+[\omega,v]-\{\phi,u\}, \nonumber\\
\delta B &=& du+\{\omega,u\}+[B,v]+[\phi,b],\\
\delta\Omega &=& db+[\omega,b]+[\Omega,v]-\{B,u\}+[\phi,b], \nonumber\\
\delta H &=& dU+\{\omega,U\}+\{\Omega,u\}+[H,v]+[B,b]+[\phi,V],  \nonumber
\end{eqnarray}
where we have introduced the following notation:
\begin{eqnarray}
  {\cal V} & = & {\mbox{\bf 1}} \hat{a} + {\mbox{\bf i}} a  \nonumber \\
          & \equiv & {\mbox{\bf 1}}(v + b + V) +  
                     {\mbox{\bf i}}(u + U),
\label{eqn:4d-gp}
\end{eqnarray}
with $v$, $u$, $b$, $U$, $V$ being 0-, 1-, 2-, 3-, 4-form bosonic 
gauge parameters, respectively.

We now introduce generalized variables completely 
parallel to the two-dimensional eqs.(\ref{eqn:Ve}) and (\ref{eqn:Vo}):
\begin{eqnarray}
 {\cal{V}}_{2n} & = & {\mbox{\bf j}} \left(u_{2n} + 
                    U_{2n}\right) 
                    + {\mbox{\bf k}} \left( v_{2n} +  
                    b_{2n} + V_{2n}\right)  
                     \  \in \Lambda_{-},                
\label{eqn:Ve4}\\
 {\cal{V}}_{2n+1} & = &  {\mbox{\bf 1}} \left( v_{2n+1} + 
                b_{2n+1} +  V_{2n+1}\right)  
               - {\mbox{\bf i}}\, \left(u_{2n+1} + 
                 U_{2n+1} \right) \  \in \Lambda_{+}, 
\label{eqn:Vo4}\\
                & &  \hspace{9cm} n = 0, 1, 2, \cdots. \nonumber
\end{eqnarray}
By applying the above definitions of generalized variables ${\cal{V}}_{2n}$ 
and ${\cal{V}}_{2n+1}$ to eqs.(\ref{eqn:ir}) and (\ref{eqn:irr}), 
we can show the infinite reducibility of 
the four-dimensional model based on the same arguments of two-dimensional 
case. 

Next we need to introduce infinite series of generalized fields 
corresponding to the generalized variables:
\begin{eqnarray}
& & C_n, \ \ C_{n\mu}, \ \ \frac{1}{2!}C_{n\mu\nu}, \ \ 
      \frac{1}{3!}C_{n\mu\nu\rho}, \ \ \frac{1}{4!}C_{n\mu\nu\rho\sigma},
\label{eqn:4d-cnc} \\
& & \hspace{8cm} n=0, \pm 1, \pm 2, \cdots, \pm\infty, \nonumber 
\end{eqnarray}
where the index $n$ indicates the ghost number of the fields and 
the fields with even (odd) ghost number are bosonic (fermionic).
The fields with ghost number 0 are the classical fields
$$
C_0=\phi, \ \ C_{0\mu}=\omega_\mu, \ \ 
  C_{0\mu\nu}=B_{\mu\nu}, \ \ 
      C_{0\mu\nu\rho}=\Omega_{\mu\nu\rho}, \ \ 
         C_{0\mu\nu\rho\sigma}=H_{\mu\nu\rho\sigma}.
$$
Then we redefine a generalized gauge field
\begin{eqnarray}
\widetilde{{\cal A}} &=& {\mbox{\bf 1}}\psi + {\mbox{\bf i}} \hat{\psi} +  
            {\mbox{\bf j}} A + {\mbox{\bf k}} \hat{A} 
            \ \in \Lambda_-, 
\label{eqn:4d-ggf} \\
&&\psi  =  \sum_{n = -\infty}^{\infty} \Big(C_{2n+1 \mu} dx^{\mu}+
                \frac{1}{3!}C_{2n+1\mu\nu\rho}
                 dx^\mu\wedge dx^\nu\wedge dx^\rho \Big),
                                                             \nonumber \\
&&\hat{\psi} =  \sum_{n = -\infty}^{\infty}  \Big( C_{2n+1} + 
             \frac{1}{2!} C_{2n+1 \mu \nu} dx^{\mu}\wedge dx^{\nu} +
             \frac{1}{4!}C_{2n+1 \mu\nu\rho\sigma}
                    dx^\mu\wedge dx^\nu\wedge dx^\rho\wedge dx^\sigma\Big),
                                                              \nonumber \\
&&A  =  \sum_{n = -\infty}^{\infty} \Big(C_{2n \mu} dx^{\mu}+
                \frac{1}{3!}C_{2n\mu\nu\rho}
                      dx^\mu\wedge dx^\nu\wedge dx^\rho \Big),
                                                             \nonumber \\
&&\hat{A} =  \sum_{n = -\infty}^{\infty}  \Big( C_{2n} + 
             \frac{1}{2!} C_{2n \mu \nu} dx^{\mu}\wedge dx^{\nu} +
             \frac{1}{4!}C_{2n \mu\nu\rho\sigma}
                      dx^\mu\wedge dx^\nu\wedge dx^\rho\wedge dx^\sigma\Big),
                                                              \nonumber 
\end{eqnarray}
where we have explicitly shown the differential form dependence.

The definition of the generalized antibracket defined in 
subsection 3.2 is universal in any even dimensions as far as 
the generalized gauge field is properly defined as an element of 
$\Lambda_-$ class.
Using the above definitions of generalized gauge fields, we claim that 
the same form of action as in two dimensions 
\begin{equation}
\displaystyle{S_{min}=
        \int  \ {\mbox{Tr}}^0_{\mbox{\bf {k}}} 
             \left( \frac {1}{2}\widetilde{{\cal A}}Q\widetilde{{\cal A}} 
                + \frac {1}{3} \widetilde{{\cal A}}^3 \right)},
\label{eqn:mina2}
\end{equation}
is the minimal part of quantized action in four dimensions. 
There is a natural procedure to derive BRST transformation, to prove 
nilpotency of BRST transformation and to show that $S_{min}$ satisfies 
master equation, by using the generalized antibracket 
arguments of subsection 3.2.
The BRST transformations and the nilpotency have the same form 
as (\ref{eqn:dl}) and (\ref{eqn:nilp})
\begin{eqnarray*}
 \delta_{\lambda} \widetilde{{\cal A}} 
 &\equiv& (\widetilde{{\cal A}},\widetilde{S})_{\lambda,{\bf k}}
   = -\widetilde{{\cal F}} \ {\mbox{\bf i}} \lambda, \\
 s^2\widetilde{{\cal A}}\lambda_2\lambda_1
 &\equiv& \delta_{\lambda_2} \delta_{\lambda_1} \widetilde{{\cal A}} 
   = 0.       
\end{eqnarray*}
We can then show that $S_{min}$ satisfies the master equation 
$$
  \delta_{\lambda} S_{min} 
    = (S_{min},S_{min})_{\lambda,{\bf k}}
    = ( S_{min} , S_{min} ) \cdot \lambda = 0,
$$
where $(~,~)$ is the original antibracket defined by 
(\ref{eqn:me}) with the following identifications of antifields:
\begin{eqnarray}
\frac{1}{4!}\epsilon^{\mu\nu\rho\sigma}C_{-2n+1 \mu\nu\rho\sigma}
                            &=&C^*_{2(n-1)}, \hspace{1.5cm}
\frac{1}{4!}\epsilon^{\mu\nu\rho\sigma}C_{-2n \mu\nu\rho\sigma}
                            =-C^*_{2n-1}, \nonumber \\
\frac{1}{3!}\epsilon^{\mu\nu\rho\sigma}C_{-2n+1 \nu\rho\sigma}
                            &=&C^{\mu*}_{2(n-1)}, \hspace{1.5cm}
\frac{1}{3!}\epsilon^{\mu\nu\rho\sigma}C_{-2n \nu\rho\sigma}
                            =C^{\mu*}_{2n-1}, \nonumber \\
\frac{1}{2!}\epsilon^{\mu\nu\rho\sigma}C_{-2n+1 \rho\sigma}
                            &=&-C^{\mu\nu*}_{2(n-1)}, \hspace{1.5cm}
\frac{1}{2!}\epsilon^{\mu\nu\rho\sigma}C_{-2n \rho\sigma}
                            =-C^{\mu\nu*}_{2n-1}, 
\label{eqn:4d-idantif} \\
\epsilon^{\mu\nu\rho\sigma}C_{-2n+1 \sigma}
                            &=&C^{\mu\nu\rho*}_{2(n-1)}, \hspace{1.5cm}
\epsilon^{\mu\nu\rho\sigma}C_{-2n \sigma}
                            =C^{\mu\nu\rho*}_{2n-1}, \nonumber \\
\epsilon^{\mu\nu\rho\sigma}C_{-2n+1}
                            &=&C^{\mu\nu\rho\sigma*}_{2(n-1)}, \hspace{1.5cm}
\epsilon^{\mu\nu\rho\sigma}C_{-2n }
                            =-C^{\mu\nu\rho\sigma*}_{2n-1}. \nonumber 
\end{eqnarray}
For completeness we give the explicit forms of the BRST transformations for 
the component fields in the appendix.

In order to fix the gauge we introduce the nonminimal action
\begin{eqnarray}
S_{nonmin}&=&\int d^4x\sum_{n=1}^{\infty}\mbox{Tr} \Bigl(
             {\bar{C}}^*_{n\mu\nu\rho}b^{\mu\nu\rho}_{n-1} 
             +{\bar{C}}^*_{n\mu\nu}b^{\mu\nu}_{n-1}
             +{\bar{C}}^*_{n\mu}b^{\mu}_{n-1}
             +{\bar{C}}^*_{n}b_{n-1} \nonumber \\
         & & \hspace{2.5cm} 
             +\eta^{\mu\nu*}_{n-1}\pi_{n\mu\nu}
             +\eta^{\mu*}_{n-1}\pi_{n\mu}
             +\eta^*_{n-1}\pi_n \nonumber \\
         & & \hspace{2.5cm} 
             +C^{*\mu}_n\rho_{n-1 \mu}
             +\xi^*_n\rho_{n-1}
             +\zeta^*_{n-1}\sigma_n \Bigr),
\label{eqn:4d-nonmin}
\end{eqnarray}
where the ghost number of nonminimal fields is $n$ for $\eta_n$, $\eta^\mu_n$, 
$\eta^{\mu\nu}_n$, $\pi_n$, $\pi_{n \mu}$ and $\zeta_n$, and $-n$ for 
${\bar{C}_n}$, ${\bar{C}^{\mu}_n}$, ${\bar{C}^{\mu\nu}_n}$, 
${\bar{C}^{\mu\nu\rho}_n}$, $b_n$, $b^{\mu}_n$, $b^{\mu\nu}_n$, 
$b^{\mu\nu\rho}_n$, $\xi_n$ and $\xi^\mu_n$, respectively.
The BRST transformations of these fields are defined by the nonminimal action, 
as usual.
In order to impose the Landau type gauge-fixing condition for 
the antisymmetric tensor fields in each sector of the ghost number, 
we adopt the gauge fermion
\begin{eqnarray}
\Psi &=& \int d^4x \sum_{n=1}^{\infty} \mbox{Tr} \Bigl( 
       {\bar{C}}^{\mu\nu\rho}_n \partial^\sigma C_{n-1 \mu\nu\rho\sigma}
       +{\bar{C}}^{\mu\nu}_n \partial^\rho C_{n-1 \mu\nu\rho}
       +{\bar{C}}^{\mu}_n \partial^\nu C_{n-1 \mu\nu}
       +{\bar{C}}_n \partial^\mu C_{n-1 \mu} \nonumber \\
    & & \hspace{2.5cm}
       +{\bar{C}}^{\mu\nu\rho}_n \partial_\mu \eta_{n-1 \nu\rho}
       +{\bar{C}}^{\mu\nu}_n \partial_\mu \eta_{n-1 \nu}
       +{\bar{C}}^{\mu}_n \partial_\mu \eta_{n-1}  \nonumber \\
    & & \hspace{2.5cm}
       +\xi^\mu_n \partial^\nu \eta_{n-1 \mu\nu}
       +\xi_n \partial^\mu \eta_{n-1 \mu}
       +\zeta_{n-1}\partial_\mu \xi^\mu_n \Bigr).
\label{eqn:4d-ps}
\end{eqnarray}
Eliminating the antifields by 
$\Phi^*_A=\frac{\partial \Psi}{\partial \Phi^A}$, 
we obtain the complete form of the four-dimensional quantized 
gauge-fixed action.

%%%%%%%%%%%
\subsection{General features in the quantization of 
            arbitrary even-dimensional models}

As we have shown in the previous subsection, the quantization procedure in two 
dimensions and four dimensions goes exactly parallel with minor modifications 
of introducing new fields in higher dimensions. 
In other words if we try to formulate the quantization procedure in terms of 
the generalized gauge fields and parameters, it is dimension independent. 
In order to stress this point we list the general procedure of quantizing 
the even-dimensional generalized Chern-Simons actions by 
Lagrangian formulation.
\begin{enumerate}
\item We first define generalized gauge field ${\cal A}$ and parameter 
${\cal V}$ of the form of (\ref{eqn:ggf}) and (\ref{eqn:ggp}) 
in terms of component fields 
explicitly as in (\ref{eqn:ggff}) and (\ref{eqn:4d-gp}). 
\item We then obtain concrete forms of, 
even-dimensional bosonic action 
$S^b_e$ of (\ref{eqn:comexp}), gauge transformations (\ref{eqn:gt-com}) 
and equations of motion (\ref{eqn:eqmotions}), in terms of component fields.
\item We introduce generalized variables ${\cal V}_{2n}$ and 
${{\cal V}_{2n+1}}$ as in (\ref{eqn:Ve}), (\ref{eqn:Vo}) and 
(\ref{eqn:Ve4}), (\ref{eqn:Vo4}). 
Then infinite reducibility is a natural consequence of the relations 
(\ref{eqn:ir}) and (\ref{eqn:irr}).   
\item Corresponding to the infinite reducibility, we introduce infinite 
series of generalized fields as in (\ref{eqn:cnc}) and (\ref{eqn:4d-cnc}) 
and redefine the generalized gauge field as in (\ref{eqn:2d-ggf}) and 
(\ref{eqn:4d-ggf}). 
\item The minimal action $S_{min}$ can be defined in terms of the generalized 
gauge fields as in (\ref{eqn:MA}) and (\ref{eqn:mina2}).  
\item The BRST transformation is given by 
$ s\widetilde{\cal{A}} = - \widetilde{\cal{F}} {\mbox{\bf{i}}} $ 
and its nilpotency 
can be shown by (\ref{eqn:nilp}).
\item $S_{min}$ satisfies master equation as in (\ref{eqn:master}) 
under the identification that the negative ghost number fields are 
identified with antifields in a proper way as in (\ref{eqn:idantif}) 
and (\ref{eqn:4d-idantif}). 
\item The nonminimal action $S_{nonmin}$ and the gauge fermion $\Psi$ 
for Landau type gauge-fixing 
can be obtained as a natural extension of two- and four- dimensional 
expressions (\ref{eqn:2d-nonmin}), (\ref{eqn:ps}) and 
(\ref{eqn:4d-nonmin}), (\ref{eqn:4d-ps}). 
\item Substituting the antifields by 
$\Phi^*_A=\frac{\partial \Psi}{\partial \Phi^A}$ into $S_{min}+S_{nonmin}$, 
we obtain the complete form of the quantized gauge-fixed action.
\end{enumerate} 

The quantization of Hamiltonian formulation in arbitrary even dimensions 
will be carried out in the similar way as in the two-dimensional case.
In general the quantizations by the Hamiltonian formulation and 
the Lagrangian formulation should give the same result. 
We have explicitly shown the equivalence in two dimensions. 
For the generalized Chern-Simons actions this point was a priori not clear 
because of the regurality violation. 
It was, however, explicitly shown that the regularity 
violating constraints 
can be used as first class constraints in the Hamiltonian formulation and 
the result coincides with that of Lagrangian formulation.

In the process of the Hamiltonian quantization we have found several 
interesting facts.
The first class constraints of Hamiltonian formulation are infinitely 
reducible and thus we need to introduce infinite series of ghosts and 
ghost momenta to quantize the system.
If we properly define a Hamiltonian version of generalized gauge 
field $\widetilde{\cal A}$ as in the two-dimensional case 
(\ref{eqn:ah}), we can define the same form of BRST differential 
(\ref{eqn:sham}) 
as that of Lagrangian formulation. 
What is further surprising is that the BRST charge has again the 
Chern-Simons form of the fermionic sector as can be seen in 
(\ref{eqn:om}).

We can generalize this result in the following: 
We take an even-dimensional bosonic action as classical action
$$
 S^b_e=\int_{M_d} \mbox{Tr}_{\bf k}
        \left( \frac{1}{2}{\cal A}Q{\cal A}+\frac{1}{3}{\cal A}^3 \right),
$$
we then obtain the quantized minimal action of Lagrangian formulation
$$
  S_{min}  =   \int_{M_d}  \ {\mbox{Tr}}^0_{\mbox{\bf {k}}} 
             \left( \frac {1}{2}\widetilde{{\cal A}}Q\widetilde{{\cal A}} 
                + \frac {1}{3} \widetilde{{\cal A}}^3 \right),
$$
where we should take ghost number zero sector.
The minimal version of BRST charge in the Hamiltonian formulation 
for the even-dimensional generalized Chern-Simons action is 
again an generalized Chern-Simons action which is an odd-dimensional 
fermionic action and should be one dimension lower than the 
minimal action $S_{min}$
$$
  \Omega_{min}  =   \int_{M_{d-1}}  \ 
             {\mbox{Tr}}^1_{\mbox{\bf {1}}} 
             \left( \frac {1}{2} \widetilde{{\cal A}}Q'\widetilde{{\cal A}} 
                + \frac {1}{3} \widetilde{{\cal A}}^3 \right), 
$$
where we should take the ghost number one sector. 
In more general cases for a graded Lie algebra, we must take Htr instead 
of the simple trace Tr in the above.

Although it is outside of the scope of this paper, we claim 
that the odd-dimensional case goes completely parallel to 
the above even-dimensional case with the replacements of the 
above expressions by
${\bf {k}}\rightarrow {\bf {j}}, {\bf {1}}\rightarrow {\bf {i}}$ 
and ${\mbox{Tr}}\rightarrow {\mbox{Str}}$, respectively.

%%%%%%%%%%%%%%%%%%%%
\section{Conclusions and discussions} 

\setcounter{equation}{0}
\setcounter{footnote}{0}

We have investigated the quantization of the even-dimensional version of 
the generalized Chern-Simons actions by the Lagrangian and Hamiltonian 
formulations. 
We have found that the models formulated by the generalized Chern-Simons 
actions are in general infinitely reducible 
and thus we need to introduce the infinite series of ghost fields 
and the quantizations turn out to be highly nontrivial. 
It is, however, an important characteristic of the generalized Chern-Simons 
formulation that the generalized gauge field 
can accommodate the infinite number of fields in a compact form and treat 
them in a unified way. 
We have then found the solution of the master equation in the Lagrangian 
formulation, the BRST invariant 
minimal action, which has again the same Chern-Simons form as the 
starting classical action. 
With careful considerations of degrees of freedom, we have 
obtained the gauge-fixed actions by 
defining gauge fermions with Landau type gauge-fixings.
In the Hamiltonian formulation we have found two first class constraints which 
break Dirac's regularity condition but lead to the same action as that of 
Lagrangian formulation under a proper choice of the gauge fermion 
and identifications of the fields. 

The quantizations of the models have been successfully carried out 
while there remains 
another question if the introduction of infinite series of the ghost fields 
does not cause any problems at the quantum level. 
The generalized Chern-Simons actions are in general expected to be 
topological since 
the classical actions have the same number of gauge fields and 
parameters and thus all the gauge fields could be gauged away.
The story is, however, not so simple because of the infinite reducibility. 
As we have shown in section 5, the infinite series of ghost fields 
cancel out the quantum effects and then the partition function becomes 
numerically one and thus the topological nature is kept even 
at the quantum level.  
In formulating these contributions from an infinite number of ghosts, 
we have used zeta function regularization. 
The correlation functions have the similar nature as the partition function 
but have some possible problem in the unphysical sector of loop contributions 
because of the infinite number of the ghost fields. 
Since the problem appears only in the unphysical sector the models are 
still expected to be consistent at the quantum level.  
There remain, however, some open problems in the unphysical sector, 
which might be related with the regularization problem.

As we have shown in section 6 the quantization procedure of the generalized 
Chern-Simons actions is dimension independent with minor 
modifications of 
introducing new higher form fields in higher dimensions.
In other words the minimal part of the quantized action have the same 
Chern-Simons 
form and the gauge fermion can be introduced with the similar forms for any 
even dimensions. 
We have already suggested that the quantization procedure in odd dimensions 
will be carried out in a parallel way except that we need to care about 
the graded Lie algebra with the supertrace Str. 
The quantizations of the odd-dimensional generalized Chern-Simons actions, 
which includes the standard three dimensional Chern-Simons action as a 
special case, will be given elsewhere~\cite{kstu2}. 
In finding dimension independent formulations the proposed generalized 
antibracket formulation was helpful to find the BRST transformation, 
its nilpotency and the solution of the master equation. 
This formulation is, however, proposed in heuristic bases and thus 
needs sound mathematical backgrounds which might propose a new aspect of 
the quantization procedure of the generalized Chern-Simons theory. 

It is interesting to consider possibly physical aspects of the introduction 
of an infinite number of the ghost fields. 
An immediate consequence is a democracy of ghosts and classical fields, 
{\it i.e.}, the classical fields are simply the zero ghost number sector 
among infinitely many ghost fields and thus the classical gauge fields 
and ghost fields have no essential difference in the minimal action.
Furthermore fermionic and bosonic gauge fields are treated in an 
equal base and the series of infinite ghosts originated from the 
classical fermionic and bosonic fields are complimentary.
In other words if we only introduce bosonic classical fields in the 
starting action we need to introduce fermionic 
fields with odd integer ghost number 
and bosonic fields with even integer ghost number as in the previous 
paper~\cite{kstu1}. 
If we introduce the classical fermionic gauge fields 
as in section 2, the odd and even nature 
should be reversed for the ghost numbers when introducing 
the corresponding ghost fields to the fermionic gauge 
fields.
It seems to mean that even the fermionic and bosonic fields have no essential 
difference in the generalized Chern-Simons theory. 
In other words fermionic fields, bosonic fields, classical fields and 
ghost fields 
are mutually inter-related via the quantization procedure.

Another surprising result which became clear after the quantization of 
Hamiltonian formulation is that the minimal part of the BRST charge of 
Hamiltonian quantization is the odd-dimensional fermionic 
counterpart of the generalized Chern-Simons action.
This is again suggesting that the BRST charge of Hamiltonian formulation  
and the minimal action of Lagraingian formulation are inter-related 
via the quantization procedure. 
In these inter-related correspondences the quaternions are again playing the 
fundamental role to relate fermions, bosons, even dimensions and 
odd dimensions. 

It may be important to note at this stage that the generalized form of 
the BRST transformation suggests a new type of differential. 
The BRST transformation 
$
 s \widetilde{{\cal A}} 
 = -\widetilde{{\cal F}} \ {\mbox{\bf i}} ,
$ 
becomes 
$
s_l \widetilde{{\cal A}} 
 = -{\mbox{\bf i}} \ \widetilde{{\cal F}} ,
$ 
when it is defined as a left variation in accordance with the exterior 
derivative. 
Then this can be written as 
$
(-{\mbox{\bf i}}s_l + {\mbox{\bf j}}d)\widetilde{{\cal A}} 
+ \widetilde{{\cal A}}^2 \ = \ 0. 
$
This is suggesting an existence of a differential into ghost direction. 
The combined differential of the new differential and the old exterior 
derivative, $\tilde{Q} = -{\bf i}s_l + {\bf j}d$, provides 
a flat connection condition which could be the equation of motion of 
a newly defined \lq\lq Chern-Simons action". 
The definition of the new differential suggests that ghost is equivalent 
to a product of the differential form and the quaternion {\bf k}, 
which is exactly the result of the previous treatment where 
the equivalence between the generalized Chern-Simons actions and 
topological particle field theory actions was shown~\cite{kw4}.

In the analyses of the quantization of the generalized Chern-Simons theory 
with abelian $gl(1,{\bf R})$ algebra, 
it was pointed out that a physical degree of freedom which 
did not exist at the classical level 
appeared in the constant part of the zero form field 
at the quantum level due to the violation of the regularity~\cite{kos}. 
This situation is unchanged even in the nonabelian cases. 
We know that the zero form field plays an important role in the generalized 
Chern-Simons theories as emphasized in the classical 
discussions~\cite{kw2,kw3}.
In particular a constant component of the zero form field played 
a role of physical order parameter between the gravity and nongravity phases 
for particular choices of nonabelian gauge algebra.
By the analyses of the quantization of Hamiltonian formulation 
it became clear that the constant mode of the zero form still remains 
as a physical mode in the quantum level. 

As we have stressed in the introduction the success of the quantization 
of the generalized Chern-Simons actions in even dimensions leads naturally 
to the quantization of topological gravities in two and four dimensions 
which were classically well defined~\cite{kw2,kw3}.

%%%%%%%%%%%%%%%%%%%%%

\vskip 1cm

\noindent{\Large{\bf Acknowledgments}}\\
This work is supported in part by Japan Society for the Promotion of 
Science under the grant number 09640330.

%%%%%%%%%%%%%%%%%%%%%

\newpage

\noindent{\Large{\bf Appendix A. \\Actions of $D$ 
and $\stackrel{(k)}{s}$ for all fields}}

\begin{eqnarray*}
 D P_{2n}^B & = & 
               [ P_{2n}^B , \eta_1^F ] + \{ P_{2n}^{'B} , \eta_1^{'F} \},\\
 D P_{2n-1}^F & = & 
          \{ P_{2n-1}^F , \eta_1^F \} + \{ P_{2n-1}^{'F} , \eta_1^{'F} \},\\
 D P_{2n}^{'B} & = & [ P_{2n}^{'B} , \eta_1^F ], \\ 
 D P_{2n-1}^{'F} & = & \{ P_{2n-1}^{'F} , \eta_1^F \}, \\
 \stackrel{(2k)}{s} \omega_1 & = & - [ P_{2k}^B , \eta_{2k+1}^F ] 
                                 - \{ P_{2k}^{'B} , \eta_{2k+1}^{'F} \}, \\ 
 \stackrel{(2k-1)}{s} \omega_1 & = & - \{ P_{2k-1}^F , \eta_{2k}^B \} 
                                 - [ P_{2k-1}^{'F} , \eta_{2k}^{'B} ], \\
 \stackrel{(2k)}{s} \pi_{\omega_1} & = & 
                               - [ \eta_{2k+1}^F , P_{2k}^{'B} ], \\ 
 \stackrel{(2k-1)}{s} \pi_{\omega_1} & = & 
                            [ \eta_{2k}^B , P_{2k-1}^{'F} ], \\
 \stackrel{(2k)}{s} \eta_{2n}^B & = & 
                           - [ \eta_{2(n+k)+1}^F , P_{2k}^{'B} ], \\  
 \stackrel{(2k-1)}{s} \eta_{2n}^B & = &
                           [ \eta_{2(n+k)}^B , P_{2k-1}^{'F} ], \\ 
 \stackrel{(2k)}{s} \eta_{2n-1}^F & = & 
                              \{ \eta_{2(n+k)}^B , P_{2k}^{'B} \}, \\
 \stackrel{(2k-1)}{s} \eta_{2n-1}^F  & = &
                              \{ \eta_{2(n+k)-1}^F , P_{2k-1}^{'F} \}, \\
 \stackrel{(2k)}{s} \eta_{2n}^{'B} & = & 
                              - [ P_{2k}^B , \eta_{2(n+k)+1}^F ] 
                              - \{ P_{2k}^{'B} , \eta_{2(n+k)+1}^{'F} \}, \\
 \stackrel{(2k-1)}{s} \eta_{2n}^{'B} & = & 
                              - \{ P_{2k-1}^F , \eta_{2(n+k)}^B \} 
                              - [ P_{2k-1}^{'F} , \eta_{2(n+k)}^{'B} ], \\
 \stackrel{(2k)}{s} \eta_{2n-1}^{'F} & = &
                                [ \eta_{2(n+k)}^B , P_{2k}^B ] 
                              + [ \eta_{2(n+k)}^{'B} , P_{2k}^{'B} ], \\
 \stackrel{(2k-1)}{s} \eta_{2n-1}^{'F} & = &
                                \{ \eta_{2(n+k)-1}^F , P_{2k-1}^F \} 
                              + \{ \eta_{2(n+k)-1}^{'F} , P_{2k-1}^{'F} \}, \\
 \stackrel{(2k)}{s} P_{2n}^B & = & [ P_{2(n+k)}^B , \eta_{2k+1}^F ] 
                                + \{ P_{2(n+k)}^{'B} , \eta_{2k+1}^{'F} \}, \\
 \stackrel{(2k-1)}{s} P_{2n}^B & = & \{ P_{2(n+k)-1}^F , \eta_{2k}^B \} 
                                + [ P_{2(n+k)-1}^{'F} , \eta_{2k}^{'B} ], \\
 \stackrel{(2k)}{s} P_{2n+1}^F & = & \{ \eta_{2k+1}^F , P_{2(n+k)+1}^F \} 
                             + \{ \eta_{2k+1}^{'F} , P_{2(n+k)+1}^{'F} \}, \\
 \stackrel{(2k-1)}{s} P_{2n+1}^F & = & [ \eta_{2k}^B , P_{2(n+k)}^B ] 
                               + [ \eta_{2k}^{'B} , P_{2(n+k)}^{'B} ], \\
 \stackrel{(2k)}{s} P_{2n}^{'B} & = & 
                              - [ \eta_{2k+1}^F , P_{2(n+k)}^{'B} ], \\ 
 \stackrel{(2k-1)}{s} P_{2n}^{'B} & = & 
                               [ \eta_{2k}^B , P_{2(n+k)-1}^{'F} ], \\ 
 \stackrel{(2k)}{s} P_{2n+1}^{'F} & = & 
                                 \{ \eta_{2k+1}^F , P_{2(n+k)+1}^{'F} \}, \\ 
 \stackrel{(2k-1)}{s} P_{2n+1}^{'F} & = & 
                                 \{ \eta_{2k}^B , P_{2(n+k)}^{'B} \}.
\end{eqnarray*}

\newpage
%%%%%%%%%%%%%%%%%%%%%
\noindent{\Large{\bf Appendix B. \\BRST transformations of the four-dimensional model}}

\begin{eqnarray*}
sC^{(0)}_{2n}&=&-\sum_{m=-\infty}^{+\infty}
              \left[C^{(0)}_{2m+1},C^{(0)}_{2(n-m)}\right], \\
sC^{(0)}_{2n-1}&=&\sum_{m=-\infty}^{+\infty}
     \left(\frac{1}{2}\left\{C^{(0)}_{2m},C^{(0)}_{2(n-m)}\right\}
       +\frac{1}{2}\left\{C^{(0)}_{2m-1},C^{(0)}_{2(n-m)+1}\right\}\right), \\
sC^{(1)}_{2n}&=&dC^{(0)}_{2n+1}+\sum_{m=-\infty}^{+\infty}
     \left(\left[C^{(1)}_{2m},C^{(0)}_{2(n-m)+1}\right]-
           \left\{C^{(1)}_{2m+1},C^{(0)}_{2(n-m)}\right\}\right), \\
sC^{(1)}_{2n-1}&=&dC^{(0)}_{2n}+\sum_{m=-\infty}^{+\infty}
     \left(\left[C^{(1)}_{2m},C^{(0)}_{2(n-m)}\right]-
           \left\{C^{(1)}_{2m-1},C^{(0)}_{2(n-m)+1}\right\}\right), \\
sC^{(2)}_{2n}&=&dC^{(1)}_{2n+1}+\sum_{m=-\infty}^{+\infty}
    \left(\left\{C^{(1)}_{2m},C^{(1)}_{2(n-m)+1}\right\}-
          \left[C^{(2)}_{2m+1},C^{(0)}_{2(n-m)}\right]-
          \left[C^{(0)}_{2m+1},C^{(2)}_{2(n-m)}\right]\right), \\
sC^{(2)}_{2n-1}&=&dC^{(1)}_{2n}+\sum_{m=-\infty}^{+\infty}
    \left(\frac{1}{2}\left\{C^{(1)}_{2m},C^{(1)}_{2(n-m)}\right\}+
          \left\{C^{(0)}_{2m},C^{(2)}_{2(n-m)}\right\}-
          \frac{1}{2}\left\{C^{(1)}_{2m-1},C^{(1)}_{2(n-m)+1}\right\} \right.\\
     & & \left. \hspace{3cm}
            + \left\{C^{(0)}_{2m-1},C^{(2)}_{2(n-m)+1}\right\}
                           \raisebox{5mm}{}\right), \\
sC^{(3)}_{2n}&=&dC^{(2)}_{2n+1}+\sum_{m=-\infty}^{+\infty}
    \left(\left[C^{(1)}_{2m},C^{(2)}_{2(n-m)}\right]+
          \left[C^{(3)}_{2m},C^{(0)}_{2(n-m)}\right]-
          \left\{C^{(1)}_{2m+1},C^{(2)}_{2(n-m)}\right\} \right. \\
     & & \left. \hspace{3cm} 
              - \left\{C^{(3)}_{2m+1},C^{(0)}_{2(n-m)}\right\}\right), \\
sC^{(3)}_{2n-1}&=&dC^{(2)}_{2n}+\sum_{m=-\infty}^{+\infty}
    \left(\left[C^{(1)}_{2m},C^{(2)}_{2(n-m)}\right]+
          \left[C^{(3)}_{2m},C^{(0)}_{2(n-m)}\right]+
          \left\{C^{(1)}_{2m-1},C^{(2)}_{2(n-m)+1}\right\} \right. \\
     & & \left. \hspace{3cm}
              +\left\{C^{(3)}_{2m-1},C^{(0)}_{2(n-m)+1}\right\}\right), \\
sC^{(4)}_{2n}&=&dC^{(3)}_{2n+1}+\sum_{m=-\infty}^{+\infty}
    \left(\left\{C^{(1)}_{2m},C^{(3)}_{2(n-m)+1}\right\}+
          \left\{C^{(3)}_{2m},C^{(1)}_{2(n-m)+1}\right\}-
          \left[C^{(0)}_{2m+1},C^{(4)}_{2(n-m)}\right] \right. \\
     & & \left. \hspace{3cm}
           - \left[C^{(2)}_{2m+1},C^{(2)}_{2(n-m)}\right] 
           - \left[C^{(4)}_{2m+1},C^{(0)}_{2(n-m)}\right] \right), \\
sC^{(4)}_{2n-1}&=&dC^{(3)}_{2n}+\sum_{m=-\infty}^{+\infty}
    \left(\left\{C^{(1)}_{2m},C^{(3)}_{2(n-m)}\right\}+
          \left\{C^{(0)}_{2m},C^{(4)}_{2(n-m)}\right\}+
          \frac{1}{2}\left\{C^{(2)}_{2m},C^{(2)}_{2(n-m)}\right\} \right. \\
     & & \left. \hspace{1cm}
           -\left\{C^{(1)}_{2m-1},C^{(3)}_{2(n-m)+1}\right\}
           +\left\{C^{(0)}_{2m-1},C^{(4)}_{2(n-m)+1}\right\}
           +\frac{1}{2}\left\{C^{(2)}_{2m-1},C^{(2)}_{2(n-m)+1}\right\}\right).
\end{eqnarray*}
Here the upper indices on $C$'s indicate the form degrees of the fields:  
$\displaystyle{C^{(2)}=\frac{1}{2}C_{\mu\nu}dx^\mu\wedge dx^\nu}$, for example.

\newpage
%%%%%%%%%%%%%%%%%%%%

\end{document}